\documentclass[twocolumn,showpacs,preprintnumbers,amsmath,amssymb]{revtex4}
\usepackage{graphicx}
\usepackage{bm} 
\usepackage{amsthm}
\usepackage[]{amsfonts, graphicx, amsmath}
\DeclareGraphicsExtensions{.eps,.ps,.eps.gz,.ps.gz,.eps.Z}

\newcommand{\ra}{\rangle}
\newcommand{\la}{\langle}
\newcommand{\tr}{{\rm Tr}}
\newcommand{\be}{\begin{equation}}
\newcommand{\ee}{\end{equation}}

\newtheorem{theorem}{Theorem}

\newenvironment{definition}[1][Definition]{\begin{trivlist}
\item[\hskip \labelsep {\bfseries #1}]}{\end{trivlist}}
\newenvironment{example}[1][Example]{\begin{trivlist}
\item[\hskip \labelsep {\bfseries #1}]}{\end{trivlist}}

\begin{document}

\title{Quantum walks on quotient graphs}
\author{\sc Hari Krovi}\email{krovi@usc.edu}
\author{\sc Todd A. Brun}\email{tbrun@usc.edu}
\affiliation{Communication Sciences Institute, University of Southern California, \\
Los Angeles, California 90089, USA}
\date{\today}

\begin{abstract}
A discrete-time quantum walk on a graph $\Gamma$ is the repeated application of a unitary evolution operator to a Hilbert space corresponding to the graph. If this unitary evolution operator has an associated group of symmetries, then for certain initial states the walk will be confined to a subspace of the original Hilbert space. Symmetries of the original graph, given by its automorphism group, can be inherited by the evolution operator. We show that a quantum walk confined to the subspace corresponding to this symmetry group can be seen as a different quantum walk on a smaller {\it quotient} graph. We give an explicit construction of the quotient graph for any subgroup $H$ of the automorphism group and illustrate it with examples. The automorphisms of the quotient graph which are inherited from the original graph are the original automorphism group modulo the subgroup $H$ used to construct it. The quotient graph is constructed by removing the symmetries of the subgroup $H$ from the original graph. We then analyze the behavior of hitting times on quotient graphs. Hitting time is the average time it takes a walk to reach a given final vertex from a given initial vertex. It has been shown in earlier work [Phys. Rev. A {\bf 74}, 042334 (2006)] that the hitting time for certain initial states of a quantum walks can be infinite, in contrast to classical random walks. We give a condition which determines whether the quotient graph has infinite hitting times given that they exist in the original graph. We apply this condition for the examples discussed and determine which quotient graphs have infinite hitting times.  All known examples of quantum walks with hitting times which are short compared to classical random walks correspond to systems with quotient graphs much smaller than the original graph; we conjecture that the existence of a small quotient graph with finite hitting times is necessary for a walk to exhibit a quantum speed-up.
\end{abstract}


\maketitle

\section{Introduction}
One of the most important goals of quantum computing is the design of fast algorithms for computational problems. The algorithms of Grover \cite{Gro96} and Shor \cite{Sho94} are among the famous examples.  These two algorithms are very different in structure:  Grover's algorithm exploits an invariant two dimensional subspace within the search space, while Shor's algorithm exploits the properties of the quantum Fourier transform. The quantum Fourier transform (QFT) has since been applied to the so called hidden subgroup problem (HSP) and has given efficient algorithms for Abelian groups and certain classes of non-Abelian groups \cite{Lom04}. But the power of the QFT in solving the non-Abelian case of the HSP maybe limited when it comes to certain non-Abelian groups such as $S_n$, the symmetric group on $n$ elements and its associated HSP-- the graph isomorphism problem. Grover's algorithm, although very useful in many search problems, gives only a quadratic speed up, and hence a straightforward application of this algorithm is not very efficient for the HSP. This is because it ignores structure in the problem which can be used to obtain a speed up. Hence, we may need new approaches to design algorithms to solve these problems. Quantum walks may provide the tools for new algorithms---first, because random walks (their classical analogues) have been very successful in the design of classical probabilistic algorithms \cite{MotRag95}; and second, because quantum walks have been shown to have properties which are useful in algorithms, such as the element distinctness problem \cite{Amb05}, and an alternative search algorithm \cite{SKW03} that has been shown to have a quadratic speed-up (the same as Grover's algorithm). In \cite{CCDFGS03}, it was shown that the quantum walk on the so-called ``glued trees'' graph reaches the final vertex from the initial vertex exponentially faster that a similar classical walk. Other algorithms based on quantum walks include matrix product verification \cite{BS06}, triangle finding \cite{MSS05} and group commutativity testing \cite{MN05}.

There are two main types of quantum walks:  continuous-time and discrete-time quantum walks. The main difference between them is that discrete time walks require a ``coin"---which is just any unitary matrix---plus an extra Hilbert space on which the coin acts, while continuous time walks do not need this extra Hilbert space. Apart from this difference, the two types of quantum walks are similar to continuous-time and discrete-time random walks in the classical case.  Discrete-time quantum walks evolve by the application of a unitary evolution operator at discrete time intervals, and continuous walks evolve under a (usually time-independent) Hamiltonian. Unlike the classical case, the extra Hilbert space for discrete-time quantum walks means that one cannot obtain the continuous quantum walk from the discrete walk by taking a limit as the time step goes to zero. The dynamics of quantum walks of both types has been studied in detail for walks on an infinite line---for the continuous-time case in  Refs.~\cite{CCDFGS03,FG98,CFG02} and for the discrete-time case in \cite{NV00,BCGJW04,BCA03a,BCA03b,BCA03c}. There has also been considerable work on other regular graphs.  The $N$-cycle is treated in \cite{AAKV00,TFMK}, and the hypercube in \cite{SKW03,MooRus02,Kem03b,KB05,KB06}.  Quantum walks on general undirected graphs are defined in \cite{Ken03, Amb03}, and on directed graphs in \cite{Mon05}. Kendon \cite{Ken06} has a recent review of the work done in this field so far, focusing mainly on decoherence. Other reviews include an introductory review by Kempe in \cite{Kem03a}, and a review from the perspective of algorithms by Ambainis in \cite{Amb03}.

Several quantities of interest have been defined for quantum walks by analogy to classical walks in \cite{AAKV01}, such as mixing time, sampling time, filling time and dispersion time. Hitting time---the average time for a particle to reach a particular final vertex---is another important quantity for classical walks on graphs. Two definitions of hitting time are given in \cite{Kem03b}, and an upper bound for one of them was found for the walk on a hypercube. A different definition of hitting time is given in \cite{KB05}, where the unitary evolution of the discrete walk is replaced by a measured walk. In such a walk, after the application of the unitary evolution operator, a measurement is performed to see if the particle is in the final vertex or not. In \cite{KB06} it was shown that graphs with sufficient symmetry can have {\it infinite} hitting times for certain initial states, a phenomenon with no classical analogue.

The idea of restricting a search to an invariant subspace of the full search space has proved very fruitful in Grover's search algorithm. In both the quantum walk-based algorithm on the hypercube in \cite{SKW03} and the ``glued-trees" graph in \cite{CCDFGS03}, the quantum algorithm works very fast by searching a smaller space, where it is known that the solution lies in this space. In this paper, we explore this concept for quantum walks on more general graphs. Using symmetry arguments, we show that it is possible to find invariant subspaces of the total Hilbert space on which the walk is defined. The automorphism group of the graph produces a group of symmetries of the evolution operator for the walk.  This group of symmetries in turn determines the invariant subspace of the walk. If the initial state is in this subspace, the quantum walk effectively evolves on a different graph---a {\it quotient} graph, which can in some cases be much smaller than the original graph. In this paper, we give a general construction of quotient graphs, given the original graph and a subgroup of its automorphism group. We determine the structure of the quantum walk on the quotient graph. We then apply the analysis of hitting times developed in \cite{KB05} and \cite{KB06} to quotient graphs, and investigate the possibility of both infinite hitting times and reduced hitting times on quotient graphs.

This paper is organized as follows. In Sec. II, we describe discrete and continuous quantum walks. In Sec. III, we define Cayley graphs and discuss their automorphisms. We give examples of a few Cayley graphs and explicitly find their automorphism groups. In Sec. IV, we describe the action of the automorphism group and show how it leads to the idea of quotient graphs. We then discuss quantum walks on quotient graphs and give examples of quotient graphs, their construction, and the effective behavior of quantum walks on these graphs. We also show how automorphisms of the quotient graph are inherited from the automorphisms of the original graph. In Sec. V, we review the definition of hitting times and then analyze the behavior of hitting times on quotient graphs. Finally, in Sec. VI, we discuss these results and present our conclusions.

\section{Quantum walks---discrete and continuous}

Quantum walks (as they have been been studied up to the present) can be either discrete-time or continuous-time, just like classical random walks. These two types of quantum walks are not exactly equivalent to the two types of classical random walks, however.  Unlike the classical case, the discrete-time quantum walk does not reduce to the continuous-time walk in a well-defined limit. Discrete-time walks need an extra Hilbert space in order to exhibit nontrivial unitary evolution; this extra space is called the ``coin''  space (from the idea that one flips a coin at each step to determine which way to walk), and taking the limit where the time step goes to zero does not eliminate the coin.  (\cite{Strauch06} offers a different treatment of this limit for the walk on the line, where it {\it is} possible to meaningfully extract the continuous-time walk as a limit of the discrete-time walk, but this has not yet been extended to more general graphs.)  Therefore, the properties of discrete and continuous walks are different. There is no obvious reason why one definition should be preferred, but in some cases it has been shown that coins make these walks faster \cite{AKR05}.

First, a bit of terminology that we will use throughout this paper.  A {\it regular} graph is one where every vertex is connected to the same number $d$ of other vertices.  This number is called the the {\it degree} of the graph.  A graph is {\it undirected} if for every edge between vertices A and B going from A to B, an edge goes from B to A as well.  In this case, we identify the edge from A to B with the edge from B to A, and consider them a single edge.  For undirected, regular graphs, we can {\it edge color} the graph by assigning numbers (or colors) to the edges at a given vertex. An edge may have different numbers assigned to it at either end, but we use the same set of numbers at every vertex.  A graph with degree $d$ such that every edge can be colored uniquely (i.e., has the same number assigned at both ends) with $d$ colors is said to be $d$-colorable.  Finally, the {\it adjacency matrix} of a graph is a matrix $A$ with elements $a_{ij}$ such that $i,j$ label vertices, and $a_{ij} = 1$ if there is an edge from $i$ to $j$ and  $a_{ij} = 0$ otherwise.  For an undirected graph the adjacency matrix is always symmetric.

\subsubsection{Discrete-time walks}

A discrete-time quantum walk can broadly be defined as the repeated application of a unitary evolution operator on a Hilbert space whose size depends on the graph. For a regular graph, this Hilbert space usually consists of the space of possible positions (i.e., the vertices) together with the space of possible directions in which the particle can move from each vertex (the coin space); for irregular graphs, this can be generalized so that there is a subspace associated with each vertex whose dimension depends on the degree of the vertex.  (In this case, however, the Hilbert space does not have a tensor product form between the coin and vertices.)  The formalism of quotient graphs developed in this article is valid for any undirected graph. Most of the graphs in the examples, however, are regular, which has been the main focus in the quantum walk literature.

We can define the Hilbert space of the walk to be $\mathcal{H}^p\otimes\mathcal{H}^c$, i.e., the tensor product of the position and direction (or coin) space. The evolution operator $\hat{U}$ is given by $\hat{U}=\hat{S}(\hat{I}\otimes\hat{C})$, where $\hat{S}$ is called the shift matrix and $\hat{C}$ is the coin matrix. The shift matrix encodes the structure of the graph and is closely related to its adjacency matrix. The vertices, numbered $|0\ra$ through $|N-1\ra$, are basis states for the vertex Hilbert space $\mathcal{H}^p$ and the set of all directions from each vertex, numbered $|1\ra$ through $|d\ra$, are basis states for the coin Hilbert space $\mathcal{H}^c$. In this basis, the shift matrix for the graph can be given the explicit form:
\[
\hat{S} = \sum_v \sum_i |v(i),j\ra\la v,i| ,
\]
where $v(i)$ is the vertex connected to $v$ along an edge which is numbered $i$ from $v$ to $v(i)$ and $j$ from $v(i)$ to $v$.

The coin matrix $\hat{C}$ acts only on the coin space, and ``flips'' the directions before the shift matrix is applied. Then $\hat{S}$ moves the particle from its present vertex to the vertex connected to it along the edge indicated by the coin direction. Though $\hat{C}$ can be any unitary matrix, usually coins with some structure are considered. The coins that we have used in our previous analysis are the Grover coin $\hat{C}_G$ and the Discrete Fourier Transform (DFT) coin $\hat{C}_D$. The matrices for these coins are given by:
\begin{equation}
\hat{C}_G = 2|\Psi\ra\la\Psi|-I = \begin{pmatrix}
  \frac{2}{d}-1 & \frac{2}{d} & \ldots & \frac{2}{d} \\
  \frac{2}{d} & \frac{2}{d}-1 & \ldots & \frac{2}{d} \\
  \vdots & \vdots & \ddots & \vdots \\
  \frac{2}{d} & \frac{2}{d} & \ldots & \frac{2}{d}-1
\end{pmatrix} ,
\label{Grover_matrix}
\end{equation}
and
\begin{equation}
\hat{C}_D=\frac{1}{\surd{d}}\begin{pmatrix}
  1 & 1 & 1 & \ldots & 1 \\
  1 & \omega & \omega^2 & \ldots & \omega^{d-1} \\
  \vdots & \vdots & \vdots & \ddots & \vdots \\
  1 & \omega^{d-1} & \omega^{2(d-1)} & \ldots &
  \omega^{(d-1)(d-1)} ,
\end{pmatrix} ,
\end{equation}
where $|\Psi\ra=\frac{1}{\surd{d}}\sum_i|i\ra$ and $\omega=\exp(2\pi i/d)$.

\subsubsection{Continuous time walks}

Continuous time quantum walks were defined by Farhi and Gutmann in \cite{FG98}. For an undirected graph $G(V,E)$, the unitary evolution operator is defined as $\hat{U}=\exp(i\hat{H}t)$, where $\hat{H}$ is obtained from the adjacency matrix of the graph. Here again, the vertices of the graph form a basis for the Hilbert space on which $\hat{U}$ is defined. This gives rise to the following Schr\"{o}dinger equation:
\begin{equation}
i\frac{d}{dt} \la v|\psi (t)\ra=\la v|\hat{H}|\psi (t) \ra.
\end{equation}
This walk has a structure very similar to that of continuous time Markov chains. $\hat{H}$ is defined as 
\begin{eqnarray}
 \label{Continuous_Ham} \hat{H}_{i,j}=\left\{
\begin{array}{cl}
   -\gamma &  \ i \neq j \mbox{ if nodes }i\mbox{ and }j\mbox{ connected} \\
         0 &  \ i \neq j \mbox{ if nodes }i\mbox{ and }j\mbox{ not connected} \\
d_i \gamma &  \ i=j
\end{array} \right.
\end{eqnarray}
where $\gamma$ is the jumping rate from a vertex to its neighbor i.e., the transitions between connected vertices happen with a probability $\gamma$ per unit time \cite{Kem03a}. But for a regular graph we can take $\hat{H}$ to be the adjacency matrix because $d_i=d$, where $d$ is the degree of the graph. This means that the Hamiltonian can be written as $\hat{H}=\gamma (D-A)$, where $D=dI$ and $A$ is the adjacency matrix of the graph. The matrix $D$ would lead to a trivial phase factor and can be dropped. $\hat{H}$ is a symmetric matrix (and hence $\hat{U}$ unitary) if the graph is undirected. Therefore, for a regular and undirected graph, the adjacency matrix $\hat{H}$, which acts as the Hamiltonian, is of the form:
\begin{equation}
H_{i,j}=\biggl\{ \begin{array}{cc} 1 & {\rm if}\ i\ {\rm and}\ j\ {\rm share\ an\ edge,} \\
0 & {\rm otherwise.} \end{array} .
\end{equation}
As can be seen, this walk has no coin and so the Hilbert space on which $\hat{U}$ acts is only the vertex space $\mathcal{H}^p$.

\section{Cayley graphs and automorphism groups}

Cayley graphs are defined in terms of a group $G$ and a set $S$ of elements from $G$, chosen such that the identity element $e\notin S$. Given $G$ and $S$, the resulting (right)-Cayley graph $\Gamma(G,S)$ is one whose vertices are labeled by the group elements and whose (edge) directions are labeled by the elements of $S$.  There is one vertex for every group element, and two vertices $g$ and $h$ are connected by a directed edge from $g$ to $h$ if $g^{-1}h\in S$, (see \cite{GroTuc87}). Another way to look at this definition is that from any vertex $g$ of a Cayley graph, there are $|S|$ outgoing edges, one to each of the vertices $gs$, $\forall s\in S$. A Cayley graph will be connected if and only if the set $S$ is a generating set for $G$, it will be undirected if $s^{-1}\in S$, $\forall s\in S$ and it will be $d$-colorable if $s^{-1}=s$, $\forall s\in S$. Cayley graphs are always regular, and the degree of a Cayley graph is $|S|$, the cardinality of the generating set.

Examples of Cayley graphs on which quantum walks have been studied include the line $\Gamma(\mathcal{Z},\{1,-1\})$; the cycle $\Gamma(\mathcal{Z}_n,\{1,-1\})$; the hypercube $\Gamma(\mathcal{Z}_2^n,X)$ where the set $X$ is the set of canonical generators $\{(1,0,0,\cdots,0),(0,1,0,\cdots,0),\dots,(0,0,0,\cdots,1)\}$; and the graph on the symmetric group $\Gamma (S_n,Y)$, where $Y$ is a generating set for $S_n$.  Let us look at the hypercube as an example of a Cayley graph where quantum walks have been extensively studied.

Consider the hypercube, which has $|\mathcal{Z}_2^n|=2^n$ vertices each with degree $|X|=n$. The vertices can be labeled by an $n$-bit string from $(0,0,\cdots,0)$ through $(1,1,\cdots,1)$. Two vertices are adjacent if they differ only by a single bit. Vertex $\vec{v}$ is connected to $n$ vertices given by $\vec{v}\oplus\vec{s}$, $\forall \vec{s}\in X$, where $\vec{v}\oplus\vec{s}$ stands for the bit-wise XOR of the bit strings $\vec{v}$ and $\vec{s}$. The unitary evolution operator for a discrete walk on the hypercube becomes $\hat{U}=\hat{S}(\hat{I}\otimes \hat{C})$, where $\hat{S}$ has the form
\[
\hat{S}=\sum_{\vec{s}}\sum_{\vec{v}}|\vec{v}\oplus\vec{s}\ra\la \vec{v}|\otimes |\vec{s}\ra\la\vec{s}|.
\]
Since the vertices of the hypercube are bit strings, and adjacent vertices are those that differ by one bit, the shift matrix of the discrete walk on the hypercube has a natural form given by
\begin{eqnarray}
\hat{S} &=& \hat{X}\otimes\hat{I}\otimes\dots\otimes\hat{I}\otimes |\vec{s_1}\ra\la\vec{s_1}|
+ \hat{I}\otimes\hat{X}\otimes\dots\otimes\hat{I}\otimes \nonumber \\
&& |\vec{s_2}\ra\la \vec{s_2}| + \ldots + \hat{I}\otimes\hat{I}\otimes\dots\otimes\hat{X}\otimes |\vec{s_n}\ra\la \vec{s_n}|,
\label{S_Pauli}
\end{eqnarray}
where $\hat{X}$ stands for the Pauli $\sigma_x$ operator. This structure of $\hat{S}$ reflects the property of the hypercube that moving along an edge from $\vec{v}$ corresponds to flipping one bit of $\vec{v}$. This structure is also useful in determining its group of symmetries as we shall see below.

An {\it automorphism} of a graph is a permutation of its vertices such that it leaves the graph unchanged. The set of all such permutations is the {\it automorphism group} of the graph. When the edge labels or colors in the graph are important, as in the case of a discrete quantum walk, we restrict ourselves to those automorphisms which preserve the edge labels. In other words, an edge connecting two vertices has the same label before and after the permutation. Such automorphisms are called {\it direction-preserving}. In general, we could consider automorphisms where we permute the direction labels along with the vertices to obtain the same graph with the same coloring. This would form a larger group $G$ of which the direction-preserving automorphisms are a subgroup $H$.

Since the vertex Hilbert space $\mathcal{H}^v$ has its basis elements in one-to-one correspondence with the vertices of the graph, and the coin Hilbert space has a basis in correspondence with the direction labels, the automorphisms (which are just permutations of vertices and directions) are permutation matrices. In fact, these are all the permutation matrices on $\mathcal{H}^v\otimes\mathcal{H}^c$ that leave $\hat{S}$ unchanged, i.e., $\{$all $\hat{P}\, |\, \hat{P}\hat{S}\hat{P}^\dag=\hat{S}$, where $\hat{P}$ is a permutation matrix$\}$. In this representation, any direction-preserving automorphism has the structure $\hat{P}_v\otimes \hat{I}_c$, where $\hat{P}_v$ acts solely on $\mathcal{H}^v$ and $\hat{I}_c$ on $\mathcal{H}^c$. Such automorphisms become important if we wish to consider the symmetries of $\hat{U} \equiv  \hat{S}(\hat{I}\otimes\hat{C})$. Clearly, any automorphism of this type is a symmetry of $\hat{U}$, since
\begin{eqnarray}
(\hat{P}_v\otimes \hat{I}_c) \left[ \hat{S}(\hat{I}\otimes\hat{C}) \right] (\hat{P}_v\otimes \hat{I}_c)^\dag =\nonumber \\
\left[ (\hat{P}_v\otimes \hat{I}_c)\hat{S}(\hat{P}_v\otimes \hat{I}_c)^\dag \right] (\hat{I}\otimes\hat{C})
= \hat{S}(\hat{I}\otimes\hat{C}) .
\end{eqnarray}
Elements of $G$ in general do not act trivially on the coin space. Because of this, they need not be symmetries of $\hat{U}$ unless the coin flip operator $\hat{C}$ respects these symmetries. 

\begin{figure}[t]
\begin{center}
\includegraphics[scale=0.5]{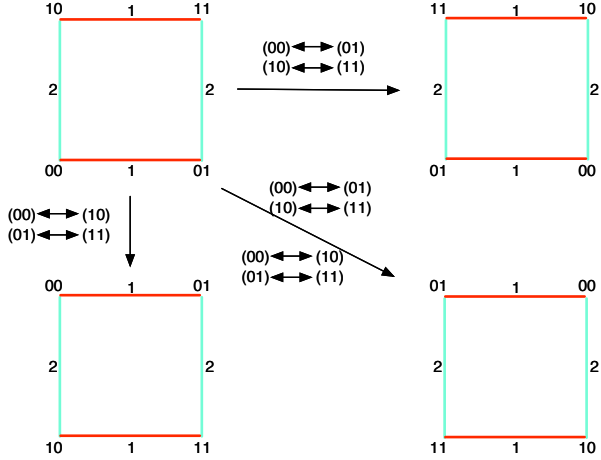}
\end{center}
\caption{(Color online)The direction-preserving automorphism group of the n=2 hypercube.} \label{fig2}
\end{figure}

To illustrate all this, consider the example of a hypercube in 2 dimensions (i.e., a square). The vertex labels are $\{(00),(01),(10),(11)\}$ (which also form a basis for $\mathcal{H}^v$); the edges connecting $(00)$ to $(01)$ and $(10)$ to $(11)$ are both labeled $1$, and the edges connecting $(00)$ to $(10)$ and $(01)$ to $(11)$ are both labeled $2$. Thus, the transformation $(00)\leftrightarrow (01)$ and $(10) \leftrightarrow (11)$, or the transformation $(00)\leftrightarrow (10)$ and $(01) \leftrightarrow (11)$, or both together, are automorphisms of this graph which need no permutation of the directions. Together with the identity automorphism (which permutes nothing), these permutations form the direction-preserving subgroup $H$. In a matrix representation on the Hilbert space $\mathcal{H}^v\otimes\mathcal{H}^c$, they are,
\[
\begin{pmatrix}
1 & 0 & 0 & 0\\
0 & 1 & 0 & 0\\
0 & 0 & 1 & 0\\
0 & 0 & 0 & 1
\end{pmatrix}\otimes \hat{I}_c ,
\begin{pmatrix}
0 & 1 & 0 & 0\\
1 & 0 & 0 & 0\\
0 & 0 & 0 & 1\\
0 & 0 & 1 & 0
\end{pmatrix}\otimes \hat{I}_c,
\]
\[
\begin{pmatrix}
0 & 0 & 0 & 1\\
0 & 0 & 1 & 0\\
0 & 1 & 0 & 0\\
1 & 0 & 0 & 0
\end{pmatrix}\otimes \hat{I}_c,
\begin{pmatrix}
0 & 0 & 1 & 0\\
0 & 0 & 0 & 1\\
1 & 0 & 0 & 0\\
0 & 1 & 0 & 0
\end{pmatrix}\otimes \hat{I}_c ,
\]
where $\hat{I}_c$ is the $2\times 2$ identity matrix acting on the coin space. These permutations can be easily seen to be (clockwise starting from top left) $H=\{\hat{I}\otimes\hat{I}\otimes\hat{I}, \hat{I}\otimes\hat{X}\otimes\hat{I}, \hat{X}\otimes\hat{I}\otimes\hat{I}, \hat{X}\otimes\hat{X}\otimes\hat{I} \}$.  Just as in the representation of $\hat{S}$ matrix in terms of the Pauli $\hat{X}$ operators given by Eq.~(\ref{S_Pauli}), this group denotes a bit flip in the first, second or both bits of each vertex, together with the identity, which gives no flip.  (See Fig.~\ref{fig2}.)

\begin{figure}[t]
\begin{center}
\includegraphics[scale=0.4]{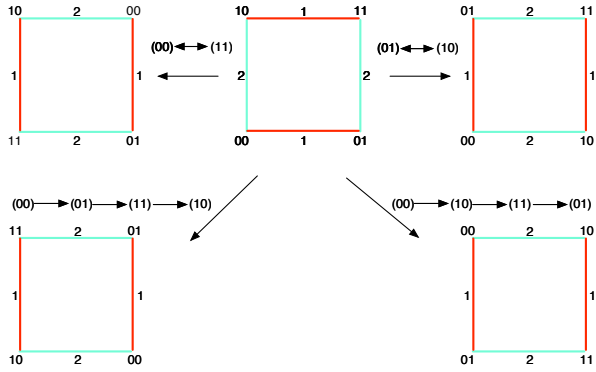}
\end{center}
\caption{(Color online) Automorphisms which interchange directions for the n=2 hypercube.} \label{fig3}
\end{figure}

The permutation $(10)\leftrightarrow (01)$, reflecting along the diagonal while keeping $(00)$ and $(11)$ fixed, will be an automorphism only if we interchange the directions $1\leftrightarrow 2$. Similarly, the permutations $(00)\leftrightarrow (11)$, $(00)\rightarrow (01)\rightarrow (11)\rightarrow (10)$ and $(00)\rightarrow (10)\rightarrow (11)\rightarrow (01)$ are automorphisms when we interchange the two directions. If we view these permutations along with those obtained above, we obtain a new group $G$ for which $H$ is a subgroup. In a matrix representation, the new automorphisms (clockwise starting from top left in Fig.~\ref{fig3}) are,
\[
\begin{pmatrix}
0 & 0 & 0 & 1\\
0 & 1 & 0 & 0\\
0 & 0 & 1 & 0\\
1 & 0 & 0 & 0
\end{pmatrix}\otimes \hat{X}_c,
\begin{pmatrix}
1 & 0 & 0 & 0\\
0 & 0 & 1 & 0\\
0 & 1 & 0 & 0\\
0 & 0 & 0 & 1
\end{pmatrix}\otimes \hat{X}_c,
\]
\[
\begin{pmatrix}
0 & 0 & 1 & 0\\
1 & 0 & 0 & 0\\
0 & 0 & 0 & 1\\
0 & 1 & 0 & 0
\end{pmatrix}\otimes \hat{X}_c,
\begin{pmatrix}
0 & 1 & 0 & 0\\
0 & 0 & 0 & 1\\
1 & 0 & 0 & 0\\
0 & 0 & 1 & 0
\end{pmatrix}\otimes \hat{X}_c,
\]
where $\hat{X}_c$ acts on the coin space and corresponds to an interchange of the two directions. These four elements of $G$ need not be symmetries of $\hat{U}$, since the coin need not be symmetric under conjugation with $\hat{X}_c$. However, for the hypercube, if we use the Grover diffusion matrix $\hat{C}_G$ as the coin, then the automorphism group $G$ is indeed the group of symmetries for the walk, since the Grover coin is symmetric under any permutation of its basis elements. The symmetry group of the evolution operator would be $H$ if the DFT coin $\hat{C}_D$ is used, since the DFT does not have permutation symmetry. Let us explicitly construct the representation of the automorphism group for the hypercube on its Hilbert space, which has $H\cong\mathcal{Z}_2^n$ and $G\cong H\cdot S_n$. In terms of the Pauli operators the representation of $H$ is $\{\hat{I}\hat{I}\hat{I}\cdots \hat{I}\otimes\hat{I}_c, \hat{X}\hat{I}\hat{I}\cdots \hat{I}\otimes\hat{I}_c, \hat{X}\hat{X}\hat{I}\cdots \hat{I}\otimes\hat{I}_c,\dots,\hat{X}\hat{X}\hat{X}\cdots \hat{X}\otimes\hat{I}_c\}$, where the tensor product symbol has been dropped in the vertex space, and $\hat{I}_c$ is the identity operator in the coin space. In fact, the representation of $H$ for any Cayley graph will be of the form $\hat{P}\otimes \hat{I}$, where $\hat{P}$ is a permutation matrix on the vertex space and $\hat{I}$ is the identity on the coin space. The group $G$ will become $H\cdot S_n=\{h \cdot \pi | h\in H, \pi\in S_n\}$, where $S_n$ is the permutation group on $n$ elements. It consists of automorphisms of the following type: any permutation of direction labels along with the permutation induced on the vertices by permuting the generators in the same way (recall that the generators are in one to one correspondence with the directions). 

It can be shown that the direction-preserving automorphism group $H$ for any Cayley graph is isomorphic to the group on which the graph is defined. This is because any direction-preserving automorphism of a Cayley graph is a left translation by a  group element, and conversely all left translations are direction-preserving automorphisms. The first part of the statement is easy to see. Consider any left translation $L_a:G\rightarrow G$ which has the action $L_a(g)=a g$, for all $g\in G$. Now, given vertices $g$ and $h$ in $G$, they are connected by an edge from $g$ to $h$ if $g^{-1}h=s$, where $s\in S$. Clearly, after the transformation we still have $(a g)^{-1}(ah)=g^{-1}h=s$ and hence this automorphism preserves the direction labels. Strictly speaking, this subgroup should be called $R(G)$, the regular representation of $G$ which is a subgroup of $S_{|G|}$ in order to clearly indicate how the automophisms permute the vertices. The other (non direction-preserving) automorphisms are not easy to find in general. Permuting the directions in some way and performing a permutation of vertices which corresponds to permuting the generators in the same way is not always an automorphism of a Cayley graph. For example, such a permutation is an automorphism of $\Gamma(S_3,\{(1,2),(2,3),(1,3)\})$ but not of $\Gamma(S_4,\{(1,2),(2,3),(3,4)\})$.

Before concluding this section, we find the automorphisms of $\Gamma(S_3,\{(1,2),(2,3),(1,3)\})$ (shown in Fig.~\ref{S_3_3gen}) since we use this as one of the examples to illustrate the idea of a quotient graph. The subgroup of direction preserving automorphisms is $R(S_3)$. Assuming that the generators $\{(1,2),(2,3),(1,3)\}=\{x_1,x_2,x_3\}$, the other automorphisms are the ones which there is any permutation of directions, say interchanging directions $x_1$ and $x_2$ and the corresponding permutation of vertices induced by this permutation of the generators i.e., $e\rightarrow e$, $x_1 \leftrightarrow x_2$, $x_1x_2\leftrightarrow x_2x_1$ and $x_1x_2x_1\rightarrow x_1x_2x_1$. There are $3!=6$ such automorphisms corresponding to each possible permutation of directions. Therefore, the automorphism group is $\mathrm{Aut}(\Gamma)\simeq R(S_3)\cdot S_3$.

\section{Quotient graphs}

\subsection{Action of an automorphism group}
Consider any undirected graph $\Gamma$ and let $V(\Gamma)$ and $E(\Gamma)$ denote its vertex and edge sets. Let the graph be colored, not necessarily consistently i.e., the edge between vertices $v_i$ and $v_j$ may be colored with a color $c_k$ in the direction from $v_i\rightarrow v_j$ and a color $c_l$ from $v_j\rightarrow v_i$. This creates the Hilbert space of positions and colors (or directions) and the total space $\mathcal{H}$ is spanned by basis vectors $|v_1,c_1\ra,\dots |v_n,c_m\ra$. Let this set of basis vectors be $X$. The set of colors at each vertex is not the same for all the vertices since the graph may be irregular. We assume that vertices having the same degree have the same set of colors. Denote by $C_v$ the set of colors used to color edges going from the vertex $v$. Thus, the shift matrix for this graph is,
\begin{equation}
S=\sum_{v_i\in V(\Gamma),k\in C_{v_i}} |v_j,c_l\ra\la v_i,c_k| .
\end{equation}
This matrix encodes the structure of the graph $\Gamma$ which includes edge colors. An automorphism of the graph $\Gamma$, as defined above is a permutation matrix which preserves $S$ under conjugation i.e., a matrix $P$ such that $PSP^\dag=S$. The set of automorphisms form a group which we denote by $G$.

Now consider a subgroup (not necessarily proper) $H$ of this automorphism group. We would like to know what kind of {\it action} this subgroup has on the graph and hence on the Hilbert space. First, we define what is meant by the term {\it action} \cite{Rot95}. 
\begin{definition}
If $X$ is a set and $G$ is a group, then $X$ is a $G${\it -set} if there is a function $\alpha : G\times X\rightarrow X$ (called a left action), denoted by $\alpha:(g,x)\rightarrow gx$, such that :
\begin{itemize}
\item $1x=x$, for all $x\in X$; and
\item $g(hx)=(gh)x$, for all $g,h \in G$ and $x\in X$ .
\end{itemize}
\end{definition}
\begin{definition}
If $X$ is a $G$-set and $x\in X$, then the $G$-{\it orbit} (or just {\it orbit}) of $x$ is
\begin{equation}
\mathcal{O}(x)=\{gx:g\in G\}\subset X .
\end{equation}
\end{definition}
The set of orbits of a $G$-set $X$ form a partition and the orbits correspond to the equivalence classes under the equivalence relation $x\equiv y$ defined by $y=gx$ for some $g\in G$. We can define the action of the subgroup $H$ of the permutation group on the set of basis elements $X$ of the Hilbert space $\mathcal{H}$ as the multiplication of its {\it matrix representation} $\sigma (H)$ (in the basis given by the vectors $X$) with a basis vector. This is a well-defined action since $\sigma (1)|x\ra=|x\ra$ and $\sigma (g) (\sigma (h) |x\ra) = (\sigma (g) \sigma (h))|x\ra = \sigma(gh) |x\ra$. Therefore, the set $X$ is partitioned into orbits under the action of $H$.

Since $H$ is a subgroup of the automorphism group, these orbits can be related to the graph $\Gamma$ through the following results.
\begin{theorem}
If $|v,c_i\ra$ and $|v,c_j\ra$ are in different orbits, then the set of all the vertices in the orbits of $|v,c_i\ra$ and $|v,c_j\ra$ are the same.
\end{theorem}
\begin{proof}
If the graph $\Gamma$ is irregular (regular graphs are just a special case), then clearly any automorphism takes a given vertex to another vertex of the same degree. Thus, automorphisms permute vertices of a certain degree among themselves. Therefore, on the Hilbert space $\mathcal{H}$, the matrix representation of any automorphism can be written as
\begin{equation}
P=\bigoplus_{d\in D} P_d ,
\end{equation}
where the set $D$ contains all the different degrees in the graph. Consider the subspace of vertices of a given degree $d$ which can be written as $\mathcal{H}_d^V\otimes \mathcal{H}_d^C$. Now, if any given permutation takes $|v_1,c_i\ra$ to $|v_2,c_j\ra$, then it takes all the basis vectors associated with $v_1$ to those of $v_2$. Thus, the set of all vertices that lie in the orbit of $|v_1,c_i\ra$ must be the same as the set of vertices that lie in the orbit of $|v_1,c_j\ra$ (if $|v_1,c_i\ra$ and $|v_1,c_j\ra$ lie in the same orbit, then this is trivially true). Since $v_1$ is arbitrary, the set of vertices in the two orbits must be the same.
\end{proof}
By an abuse of language, say that a vector $|v_1,c_1\ra$ is ``connected" to $|v_2,c_2\ra$ if the edge colored $c_1$ from vertex $v_1$ on the graph is connected to vertex $v_2$ along the color $c_2$ (i.e., the term $|v_2,c_2\ra\la v_1,c_1|$ occurs in $S$).
\begin{theorem}
If $|v_1,c_1\ra$ and $|v_2,c_2\ra$ are ``connected," $|v_1,c_1\ra$ lies in the orbit $\mathcal{O}_1$ and $|v_2,c_2\ra$ lies in orbit $\mathcal{O}_2$ (not necessarily distinct from $\mathcal{O}_1$), then each of the remaining vectors of $\mathcal{O}_1$ are ``connected" to some vector  of $\mathcal{O}_2$.
\end{theorem}
\begin{proof}
If $|v_1,c_1\ra$ and $|v_2,c_2\ra$ are connected, then there is a term of the type $|v_2,c_2\ra\la v_1,c_1|$ in $S$. When we conjugate by some automorphism $h\in H$ i.e., perform $\sigma(h)S\sigma(h)^T$, then this term transforms to $\sigma(h)|v_2,c_2\ra\la v_1,c_1|\sigma(h)^T$ and this must be a term in $S$ because $\sigma(h)S\sigma(h)^T=S$. This means that the vector that $|v_1,c_1\ra$ gets taken to, is connected to the vector that $|v_2,c_2\ra$ gets taken to by $\sigma (h)$. Since this is true for all $h\in H$, all the vectors in the orbit of $|v_1,c_1\ra$ are ``connected" to some term in the orbit of $|v_2,c_2\ra$.
\end{proof}
The above result applies equally well to any vector in $\mathcal{O}_2$. Therefore, one can think of the orbits $\mathcal{O}_1$ and $\mathcal{O}_2$ as being ``connected".

\subsection{Quotient graphs and quantum walks}
Based on the action on a graph $\Gamma$ of the subgroup $H$ of its automorphism group, consider the following construction of a graph---a {\bf quotient} graph. The set of vertices occuring in an orbit $\mathcal{O}$ is a single vertex $v_\mathcal{O}$ on the new graph and the number of orbits that have the same set of vertices is the degree of this new vertex. Thus, for a given vertex $v_\mathcal{O}$, the set of directions are the various orbits which correspond to the same vertex set. If an orbit $\mathcal{O}_1$ is ``connected" to $\mathcal{O}_2$, then the vertices $v_{\mathcal{O}_1}$ and $v_{\mathcal{O}_2}$ are connected in the quotient graph. If $\mathcal{O}_1$ and $\mathcal{O}_2$ are identical, this corresponds to a self loop for $v_{\mathcal{O}_1}$. This means that there can be self loops in the quotient graph even if there are none in the original graph. We denote the quotient graph obtained by the action of the subgroup $H$ on $\Gamma$ as $\Gamma /H$ or $\Gamma_H$.

Now consider a basis vector $|x\ra\equiv|v,c\ra$ and its $H$-orbit $\mathcal{O}_x=\{\sigma (h)|x\ra : h\in H\}$.  The vector $|\tilde{x}\ra\equiv\frac{1}{\sqrt{|\mathcal{O}_x|}}\sum_{h\in H} \sigma(h) |x\ra$ is an eigenvector of eigenvalue 1 of all the matrices $\sigma (h)$ for $h\in H$ since 
\begin{eqnarray}
\sigma(h)|\tilde{x}\ra &=& \frac{1}{\sqrt{|\mathcal{O}_x|}}\sum_{h'\in H} \sigma(h)\sigma(h')|x\ra \nonumber \\
&=& \frac{1}{\sqrt{|\mathcal{O}_x|}}\sum_{h'\in H} \sigma(hh')|x\ra \nonumber \\
&=& \frac{1}{\sqrt{|\mathcal{O}_x|}}\sum_{h''\in H} \sigma(h'')|x\ra \nonumber \\
&=& |\tilde{x}\ra .
\end{eqnarray}
Similarly, the vector $|\tilde{y}\ra$ formed from a vector $|y\ra$ of another orbit is also an eigenvector of eigenvalue 1. Each of these vectors $\{|\tilde{x}\ra\}$ are orthonormal, since they are formed from orbits and distinct orbits do not intersect and they span the simulataneous eigenspace of eigenvalue 1 of the matrices $\sigma(H)$. We denote the Hilbert space spanned by these vectors by $\mathcal{H} /H$ or $\mathcal{H}_H$.

Note that the vectors $|x\ra$ are just representatives, and any vector in its orbit could be used to generate $|\tilde{x}\ra$. Since the $\{|\tilde{x}\ra\}$ are in one to one correspondence with the orbits, we let $|\mathcal{O}_x\ra$ denote a vector in $\mathcal{H}$ and $|\tilde{x}\ra$ denote the corresponding basis vector in $\mathcal{H}_H$. 

Each basis vector in this space corresponds to a vertex and direction on the quotient graph, just as the basis vectors of $\mathcal{H}$, namely $\{|v,c\ra\}$ represent a vertex and direction on $\Gamma$. Suppose that a vertex $\tilde{v}$ on $\Gamma_H$ comes from the set of vertices in the orbit $\mathcal{O}_1$ and that the orbits $\mathcal{O}_2,\dots ,\mathcal{O}_k$ are all the orbits with the same set of vertices. Since each of these orbits is ``connected" to some other orbit (either in this set or outside), the degree of $\tilde{v}$ is $k$. Therefore, all the basis vectors $|\mathcal{O}_1\ra,\dots ,|\mathcal{O}_k\ra$ can be associated with $\tilde{v}$ and the edges along which they are  ``connected" to other orbits, as the different directions. An alternate labelling of these vectors could be $|\tilde{v},c_1\ra,\dots ,|\tilde{v},c_k\ra$ and likewise for each vertex. Note that this does not produce any natural coloring scheme induced from $\Gamma$, on the edges of $\Gamma_H$. 

We now show that any discrete quantum walk on $\Gamma$ induces a discrete quantum walk on $\Gamma_H$ as long as $\hat{U}$ respects $H$ i.e., $\sigma(h)\hat{U}\sigma(h)^\dag=\hat{U}$ $\forall h\in H$. Let us define a discrete quantum walk as the application of any unitary $\hat{U}$ which takes a particle on a given vertex $v$ to some superposition of vertices that it is connected to and the directions of only those edges which connect them to $v$. On a basis state it acts as,
\begin{equation}\label{Qwalk}
\hat{U}|v,c_i\ra=\sum_j a_j|v(c_j),c'_j\ra ,
\end{equation}
where $|v(c_j),c'_j\ra$ and $|v,c_j\ra$ are ``connected", $\sum_j |a_j|^2=1,$ $\forall j$ and the sum runs over all the colors of the edges on the side of $v$. Given this definition for a walk, we have the following results.
\begin{theorem}
Let $H$ be a subgroup of the automorphism group of $\Gamma$ and let $\hat{U}$ be a discrete quantum walk defined on $\Gamma$ such that $\hat{U}$ respects the symmetries of the subgroup i.e., $[\hat{U},\sigma(h)]=0$, $\forall h\in H$. If the initial state lies in the subspace spanned by all the orbits $\{|\mathcal{O}_i\ra\}$ under the action of $H$, then the walk is contained in the subspace.
\end{theorem}
\begin{proof}
We have,
\begin{equation}
\hat{U}^t |\mathcal{O}_i\ra=\hat{U}^t \sigma(h) |\mathcal{O}_i\ra = \sigma(h)\hat{U}^t |\mathcal{O}_i\ra .
\end{equation}
This shows that since $|\mathcal{O}_i\ra$ lies in the eigenspace of eigenvalue 1, $\hat{U}^t |\mathcal{O}_i\ra$ also lies in the same space, which is spanned by $\{|\mathcal{O}_i\ra\}$.
\end{proof}
\begin{theorem}
Let $H$ be a subgroup of the automorphism group of $\Gamma$ and let $\hat{U}$ be a discrete quantum walk defined on $\Gamma$ such that $\hat{U}$ respects the symmetries of the subgroup i.e., $[\hat{U},\sigma(h)]=0$, $\forall h\in H$. If the initial state lies in the subspace spanned by all the $H$-orbits $\{|\mathcal{O}_i\ra\}$, then $\hat{U}$ induces a walk on $\Gamma_H$ in the Hilbert space $\mathcal{H}_H$.
\end{theorem}
\begin{proof}
In order to show that $\hat{U}$ induces a walk on $\Gamma_H$, we need to show that its action is similar to Eq.~(\ref{Qwalk}):
\begin{equation}
\hat{U}|\mathcal{O}\ra = \sum_j b_j |\mathcal{O}_j\ra ,
\end{equation}
where $|\mathcal{O}_j\ra$ are the orbits connected to $|\mathcal{O}\ra$. But this follows from the fact that  if the walk moves the particle from a vector to vectors ``connected" to it, then it does the same for superpostions of vectors or the orbit states $|\mathcal{O}\ra$.
\end{proof}

We can derive the structure of this induced walk from the original walk by making use of its action on the orbit states. The induced walk on the subspace $\mathcal{H}_H$ becomes $\hat{U}_H=\sum_{\tilde{x},\tilde{y}} \la \mathcal{O}_y |\hat{U}|\mathcal{O}_x\ra | \tilde{y}\ra\la \tilde{x}|$. This defines a unitary operator in $\mathcal{H}_H$ because,
\begin{eqnarray}
\hat{U}_H^\dag\hat{U}_H &=& \sum_{\tilde{x},\tilde{y},\tilde{y}'} \la \mathcal{O}_{y'} | \hat{U}^\dag | \mathcal{O}_x\ra\la \mathcal{O}_x| \hat{U} |\mathcal{O}_y\ra |\tilde{y}'\ra\la \tilde{y}| \nonumber \\
&=& \sum_{\tilde{x},\tilde{y},\tilde{y}'} \la \mathcal{O}_{y'} | \hat{U}^\dag P_H \hat{U}|\mathcal{O}_y\ra |\tilde{y}'\ra\la \tilde{y}| \nonumber \\
&=& I_H ,
\end{eqnarray}
since $\hat{U}$ commutes with $P_H$, where $P_H$ is the projector onto $\mathcal{H}_H$. Now consider the shift matrix of the walk. Its action on $\mathcal{H}_H$ is given by $\hat{S}_H=\sum_{\tilde{x},\tilde{y}} \la \mathcal{O}_y | \hat{S} | \mathcal{O}_x\ra |\tilde{y}\ra\la \tilde{x} |$. The expression $\la \mathcal{O}_y | \hat{S} | \mathcal{O}_x\ra$ is non-zero if and only if the two orbits are ``connected". If two orbits are connected then they must be a superposition of the same number of vectors i.e., $|\mathcal{O}_x|=|\mathcal{O}_y|$ and each vector in the superposition in $|\mathcal{O}_x\ra$ is connected to one vector in the superposition in $|\mathcal{O}_y\ra$. Therefore, 
\[
\la \mathcal{O}_y | \hat{S} | \mathcal{O}_x\ra= |\mathcal{O}_x|/ \sqrt{|\mathcal{O}_x||\mathcal{O}_y}| . 
\]
Thus, 
\begin{equation}
\hat{S}_H=\sum_{\tilde{x},\tilde{y}}|\tilde{y}\ra\la \tilde{x} | .
\end{equation}
This means that the action of $\hat{S}_H$ is very similar to the action of $\hat{S}$ in that it takes the walker from any vertex to the vertex it is connected to in the quotient graph. The action of the coin which was $\hat{I}\otimes\hat{C}$ on the original graph becomes $\hat{C}_H$ on the quotient graph so that $\hat{U}_H=\hat{S}_H\hat{C}_H$. Moreover, $\hat{C}_H$ can be decomposed as follows,
\be
\hat{C}_H=\hat{C}_1\oplus\hat{C}_2\oplus \dots\oplus \hat{C}_N ,
\ee
where $N$ is the total number of vertices of the quotient graph and each $\hat{C}_i$ acts only on the basis vectors associated with the vertex $v_i$ of the quotient graph and each $\hat{C}_i$ has a dimension $d_i$ which corresponds to the degree of the $v_i$. In the following examples, such a decomposition is provided along with a list of the basis vectors on the quotient graph such that $\hat{C}_1$ acts on the first $d_1$ basis vectors, $\hat{C}_2$ acts on the next $d_2$ vectors etc.

\subsection{Examples of quotient graphs}
In this section, we illustrate the above abstract formalism with some examples. In all of the examples we use the following notation to describe the subgroups used to find quotient graphs. The elements of the subgroups denote permutations of directions, but it is to be understood that this has to be done along with an appropriate permutation of vertices, which makes it an automorphism of the graph. Although such a permutation of vertices need not exist for every permutation of directions, they exist for the examples that we consider here. Moreover, this permutation of vertices can be specified simply:  permute the generators which are in one-to-one correspondence with the directions in the same way as the directions and this induces a permutation of vertices.

For example, let $(1,2)$ be a group element. This is the automorphism obtained by interchanging  directions $1$ and $2$ and interchanging generators $t_1$ and $t_2$ so that vertices such as $t_1t_2$ go to $t_2t_1$ etc. We do not consider direction preserving automorphisms in the following examples, since they tend to give rise to quotient graphs with self loops. Finally, we use cycle notation to denote permutations, i.e., $(1,2,3)$ means $1$ goes to $2$, $2$ goes to $3$ and $3$ goes to $1$.

\begin{figure}[h]
\begin{center}
\includegraphics[scale=0.5]{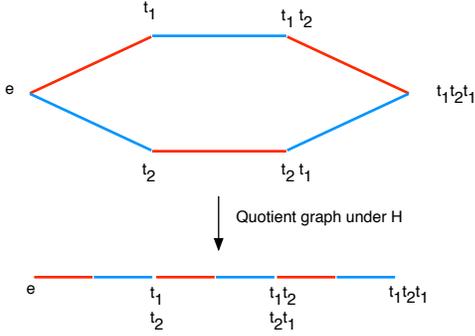}
\end{center}
\caption{(Color online) The graph $\Gamma(S_3,\{(1,2),(2,3)\})$ and its quotient graph.} \label{S_3_2gen}
\end{figure}

\begin{example}[Example 1.]
As the first example, consider the Cayley graph $\Gamma(S_3,\{(1,2),(2,3)\})$, and let $\{t_1,t_2\}=\{(1,2),(2,3)\}$. The basis vectors of the Hilbert space of the walk are $\{|e,1\ra,|e,2\ra,|t_1,1\ra ,\dots ,|t_1t_2t_1,2\ra\}$. The automorphism group of this graph is $\text{Aut}(\Gamma(S_3,T))\simeq \text{R}(S_3)Z_2$. Consider the subgroup $H=Z_2$ which corresponds to interchanging the directions $1$ and $2$. The orbits under the action of this subgroup are,
\begin{eqnarray*}
|\mathcal{O}_1\ra &=& (1/\sqrt{2}) (|e,1\ra + |e,2\ra),  \\
|\mathcal{O}_2\ra &=& (1/\sqrt{2}) (|t_1,1\ra + |t_2,2\ra),  \\
|\mathcal{O}_3\ra &=& (1/\sqrt{2}) (|t_1,2\ra + |t_2,1\ra),  \\
|\mathcal{O}_4\ra &=& (1/\sqrt{2}) (|t_1t_2,2\ra + |t_2t_1,1\ra),  \\
|\mathcal{O}_5\ra &=& (1/\sqrt{2}) (|t_1t_2,1\ra + |t_2t_1,2\ra),  \\
|\mathcal{O}_6\ra &=& (1/\sqrt{2}) (|t_1t_2t_1,1\ra + |t_1t_2t_1,2\ra). 
\end{eqnarray*}
The original and the quotient graph in this case are shown in Fig. ~\ref{S_3_2gen}. The unitary describing the quantum walk on $\Gamma$ is given by $\hat{U}=\hat{S}(\hat{I}\otimes \hat{C})$ where $\hat{S}=\hat{S}'+\hat{S}'^\dag$ and
\begin{eqnarray}
\hat{S}'=|e,1\ra\la t_1,1| + |e,2\ra\la t_2,2| + |t_1,2\ra\la t_1t_2,2| + \nonumber \\
|t_2,1\ra\la t_2t_1,1| + |t_1t_2,1\ra\la t_1t_2t_1,1| + |t_2t_1,2\ra\la t_1t_2t_1,2| . \nonumber
\end{eqnarray}
This becomes $\hat{S}_H$ on the quotient graph and is given by $\hat{S}_H=\hat{S}'_H+\hat{S}'^\dag_H$ and
\begin{equation}
\hat{S}'_H=|\mathcal{O}_1\ra\la \mathcal{O}_2| + |\mathcal{O}_3\ra\la \mathcal{O}_4| + |\mathcal{O}_5\ra\la \mathcal{O}_6| .
\end{equation}
This can also be written by giving new labels to the vertices and directions of the quotient graph,
\begin{equation}
\hat{S}'_H=|v_1,R\ra\la v_2,L| + |v_2,R\ra\la v_3,L| + |v_3,R\ra\la v_4,L| ,
\end{equation}
where we have relabeled $|\mathcal{O}_1\ra$ through $|\mathcal{O}_6\ra$ as $|v_1,R\ra$ through $|v_4,L\ra$. Note that there is no $|v_1,L\ra$ and $|v_4,R\ra$ which exactly corresponds to the way these vertices are connected in the quotient graph. Now, if we take the coin to be $C=\sigma_x$, the Pauli $X$ operator (which is also the Grover coin in two dimensions), then on the quotient graph the coin flip matrix $\hat{F}_H=(\hat{I}\otimes \hat{C})_H$ becomes,
\begin{equation}
\hat{F}_H=|v_1,R\ra\la v_1,R| + |v_4,L\ra\la v_4,L| + \hat{F}'_H +\hat{F}'^\dag_H,
\end{equation}
where $\hat{F}'_H=|v_2,L\ra\la v_2,R| + |v_3,L\ra\la v_3,R|$. It can also be written as,
\be
\hat{F}_H=\hat{1}\oplus\hat{C}'\oplus\hat{C}'\oplus\hat{1},
\ee
where $\hat{C}'=\hat{X}$, the Pauli $\sigma_x$ operator. Thus, the walk becomes $\hat{U}_H=\hat{S}_H\hat{F}_H$ i.e.,
\[
\hat{U}_H=\begin{pmatrix}
0 & 0 & 1 & 0 & 0 & 0\\
1 & 0 & 0 & 0 & 0 & 0\\
0 & 0 & 0 & 0 & 1 & 0\\
0 & 1 & 0 & 0 & 0 & 0\\
0 & 0 & 0 & 0 & 0 & 1\\
0 & 0 & 0 & 1 & 0 & 0
\end{pmatrix} .
\]
\end{example}

\begin{example}[Example 2.]
Now consider the Cayley graph $\Gamma (S_3,\{(1,2),(2,3),(1,3)\})$ where $\{(1,2),(2,3),(1,3)\}=\{t_1,t_2,t_3\}$. A subgroup of its automorphism group is $S_3$ which consists of all permutations of the three directions. Consider a subgroup of this consisting of $H_1=\{e,(1,2,3),(1,3,2)\}$. Under the action of this subgroup, the orbits are,
\begin{eqnarray*}
|\mathcal{O}_1\ra &=& (1/\sqrt{3}) (|e,1\ra + |e,2\ra + |e,3\ra),  \\
|\mathcal{O}_2\ra &=& (1/\sqrt{3}) (|t_1,1\ra + |t_2,2\ra + |t_3,3\ra),  \\
|\mathcal{O}_3\ra &=& (1/\sqrt{3}) (|t_1,3\ra + |t_2,1\ra + |t_3,2\ra),  \\
|\mathcal{O}_4\ra &=& (1/\sqrt{3}) (|t_1,2\ra + |t_2,3\ra + |t_3,1\ra),  \\
|\mathcal{O}_5\ra &=& (1/\sqrt{3}) (|t_1t_2,1\ra + |t_1t_2,2\ra + |t_1t_2,3\ra),  \\
|\mathcal{O}_6\ra &=& (1/\sqrt{3}) (|t_2t_1,1\ra + |t_2t_1,2\ra + |t_2t_1,3\ra). 
\end{eqnarray*}
The shift matrix for this walk becomes $\hat{S}_{H_1}=\hat{S}'_{H_1}+\hat{S}'^\dag_{H_1}$, where
\begin{equation}
\hat{S}'_{H_1}=|\mathcal{O}_1\ra\la \mathcal{O}_2| + |\mathcal{O}_6\ra\la \mathcal{O}_3| + |\mathcal{O}_5\ra\la\mathcal{O}_4| .
\end{equation}
We can relabel the quotient graph as shown in Fig (\ref{S_3_3gen}). The matrix $\hat{S}'_{H_1}$ becomes
\begin{equation}
\hat{S}'_{H_1}=|v_1,1\ra\la v_2,1| + |v_3,1\ra\la v_2,2| + |v_4,1\ra\la v_2,3| .
\end{equation}
If we choose the Grover coin for the walk, the walk on the quotient graph becomes
\begin{equation}
\hat{U}_{H_1}=
\begin{pmatrix}
0 & -\frac{1}{3} & \frac{2}{3} & \frac{2}{3} & 0 & 0\\
1 & 0 & 0 & 0 & 0 & 0\\
0 & 0 & 0 & 0 & 0 & 1\\
0 & 0 & 0 & 0 & 1 & 0\\
0 & \frac{2}{3} & \frac{2}{3} & -\frac{1}{3} & 0 & 0\\
0 & \frac{2}{3} & -\frac{1}{3} & \frac{2}{3} & 0 & 0
\end{pmatrix}
\end{equation}
Fig.~\ref{S_3_3gen} also shows the quotient graphs for the above Cayley graph with subgroups $H_2\simeq S_3$ and $H_3\simeq \{e,(2,3)\}$. The basis states of the quotient Hilbert space $\mathcal{H}_{H_2}$ are
\begin{eqnarray*}
|\mathcal{O}_1\ra &=& (|e,1\ra+|e,2\ra+|e,3\ra)/\sqrt{3} ,\\
|\mathcal{O}_2\ra &=& (|t_1,1\ra+|t_2,2\ra+|t_3,3\ra)/\sqrt{3},\\
|\mathcal{O}_3\ra &=& (|t_1,2\ra+|t_1,3\ra+|t_2,1\ra+|t_2,3\ra \\
&+& |t_3,1\ra+|t_3,3\ra)/\sqrt{6},\\
|\mathcal{O}_4\ra &=& (|t_1t_2,1\ra+|t_1t_2,2\ra+|t_1t_2,3\ra + |t_2t_1,1\ra \\
&+& |t_2t_1,2\ra+|t_2t_1,3\ra)/\sqrt{6},
\end{eqnarray*}
and the basis states of $\mathcal{H}_{H_3}$ are
\begin{eqnarray*}
|\mathcal{O}_1\ra &=& |e,1\ra, \\
|\mathcal{O}_2\ra &=& (|e,2\ra+|e,3\ra)/\sqrt{2}, \\
|\mathcal{O}_3\ra &=& |t_1,1\ra, \\
|\mathcal{O}_4\ra &=& (|t_1,2\ra+|t_1,3\ra)/\sqrt{2}, \\
|\mathcal{O}_5\ra &=& (|t_2,2\ra+|t_3,3\ra)/\sqrt{2}, \\
|\mathcal{O}_6\ra &=& (|t_2,3\ra+|t_3,2\ra)/\sqrt{2}, \\
|\mathcal{O}_7\ra &=& (|t_2,1\ra+|t_3,1\ra)/\sqrt{2}, \\
|\mathcal{O}_8\ra &=& (|t_1t_2,3\ra+|t_2t_1,2\ra)/\sqrt{2}, \\
|\mathcal{O}_9\ra &=& (|t_1t_2,1\ra+|t_2t_1,1\ra)/\sqrt{2}, \\
|\mathcal{O}_{10}\ra &=& (|t_1t_2,2\ra+|t_2t_1,3\ra)/\sqrt{2}.
\end{eqnarray*}

The unitary corresponding to the walk on the quotient graph of $H_2$ is,
\begin{equation}
\hat{U}_{H_2}=
\begin{pmatrix}
0 & -1/3 & 2\sqrt{2}/3 & 0\\
1 & 0 & 0 & 0\\
0 & 0 & 0 & 1\\
0 & -2\sqrt{2}/3 & 1/3 & 0
\end{pmatrix},
\end{equation}
and the one on the quotient graph of $H_3$ is,
\be
\hat{U}_{H_3}=\hat{S}_{H_3}\cdot \hat{C}_{H_3} .
\ee
The matrices $\hat{S}_{H_3}$ and $\hat{C}_{H_3}$ are given by, $\hat{S}_{H_3}=\hat{S}'+\hat{S}'^\dag$ and $\hat{C}_{H_3}=\hat{C}'\oplus\hat{C}'\oplus \hat{C}''\oplus\hat{C}''$ where,
\begin{eqnarray*}
\hat{S}'&= & (|\mathcal{O}_1\ra\la\mathcal{O}_3| + |\mathcal{O}_2\ra\la\mathcal{O}_5| + |\mathcal{O}_4\ra\la\mathcal{O}_{10}| + |\mathcal{O}_6\ra\la\mathcal{O}_9| \\
&+& |\mathcal{O}_7\ra\la\mathcal{O}_8|,
\end{eqnarray*}
\[
\hat{C}'=\begin{pmatrix}-1/3 & 2\sqrt{2}/3\\ 2\sqrt{2}/3 & 1/3\end{pmatrix} 
\]
and
\[
\hat{C}''=\begin{pmatrix}-1/3 & 2/3 & 2/3 \\ 2/3 & -1/3 & 2/3 \\2/3 & 2/3 & -1/3\end{pmatrix} .
\]
\begin{figure}[h]
\begin{center}
\includegraphics[scale=0.4]{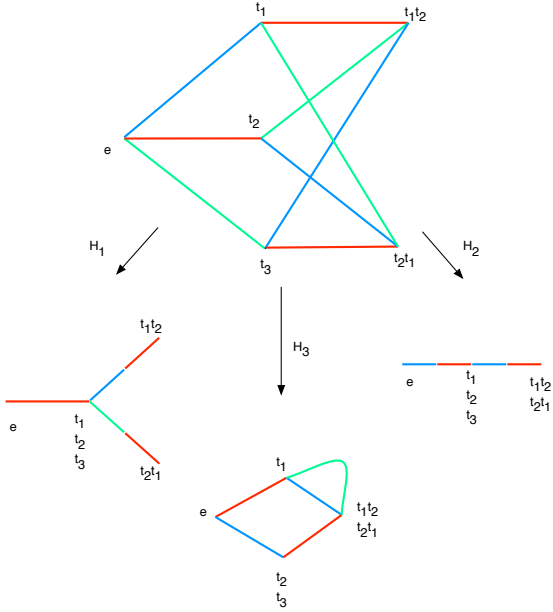}
\end{center}
\caption{(Color online) The graph $\Gamma(S_3,\{(1,2),(1,3),(2,3)\})$ and its quotient graphs.} \label{S_3_3gen}
\end{figure}
\end{example}

\begin{example}[Example 3.]
In this example, we determine the quotient graph of $\Gamma(S_4,T)$ for $T=\{(1,2),(1,3),(1,4)\}=\{t_1,t_2,t_3\}$ under the subgroup $H\simeq S_3$ which corresponds to all possible permutations of the directions at each vertex. The original and the quotient graphs are shown in Fig (\ref{S_4_3gen}), where ``$t$" has been dropped in the vertex labels. There are 16 orbits under the action of this subgroup. These are
\begin{eqnarray*}
|\mathcal{O}_1\ra &=& (|e,1\ra + |e,2\ra+|e,3\ra)/\sqrt{3},\\
|\mathcal{O}_2\ra &=& (|t_1,1\ra + |t_2,2\ra+|t_3,3\ra)/\sqrt{3},\\
|\mathcal{O}_3\ra &=& (|t_1,2\ra + |t_2,1\ra+|t_3,2\ra \\
&+& |t_2,3\ra+ |t_1,3\ra+|t_3,1\ra)/\sqrt{6},\\
|\mathcal{O}_4\ra &=& (|t_1t_2,1\ra + |t_2t_1,2\ra+|t_3t_2,3\ra \\
&+ & |t_2t_3,2\ra+ |t_3t_1,3\ra+|t_1t_3,1\ra)/\sqrt{6},\\
|\mathcal{O}_5\ra &=& (|t_1t_2,2\ra + |t_2t_1,1\ra+|t_3t_2,2\ra \\
&+& |t_2t_3,3\ra+ |t_3t_1,1\ra+|t_1t_3,3\ra)/\sqrt{6},\\
|\mathcal{O}_6\ra &=& (|t_1t_2,3\ra + |t_2t_1,3\ra+|t_3t_2,1\ra \\ 
&+& |t_2t_3,1\ra+ |t_3t_1,2\ra+|t_1t_3,2\ra)/\sqrt{6},\\
|\mathcal{O}_7\ra &=& (|t_1t_2t_1,1\ra + |t_1t_2t_1,2\ra+|t_3t_2t_3,3\ra \\ 
&+& |t_3t_2t_3,2\ra+ |t_1t_3t_1,3\ra+|t_1t_3t_1,1\ra)/\sqrt{6},\\
|\mathcal{O}_8\ra &=& (|t_1t_2t_1,3\ra + |t_3t_2t_3,1\ra+|t_1t_3t_1,2\ra)/\sqrt{3},\\
|\mathcal{O}_9\ra &=& (|t_1t_2t_3,1\ra + |t_2t_1t_3,2\ra+|t_3t_2t_1,3\ra \\
&+& |t_1t_3t_2,1\ra+ |t_2t_3t_1,2\ra+|t_3t_1t_2,3\ra)/\sqrt{6},\\
|\mathcal{O}_{10}\ra &=& (|t_1t_2t_3,2\ra + |t_1t_3t_2,3\ra+|t_2t_1t_3,1\ra \\
&+& |t_2t_3t_1,3\ra+ |t_3t_1t_2,1\ra+|t_3t_2t_1,2\ra)/\sqrt{6},\\
|\mathcal{O}_{11}\ra &=& (|t_1t_2t_3,3\ra + |t_1t_3t_2,2\ra+|t_2t_1t_3,3\ra \\
&+& |t_2t_3t_1,1\ra+ |t_3t_1t_2,2\ra+|t_3t_2t_1,1\ra)/\sqrt{6},\\
|\mathcal{O}_{12}\ra &=& (|t_3t_1t_2t_1,1\ra + |t_2t_3t_2t_1,3\ra+|t_3t_1t_2t_1,2\ra \\
&+& |t_2t_3t_2t_1,2\ra+ |t_1t_3t_1t_2,1\ra+|t_1t_3t_1t_2,3\ra)/\sqrt{6},\\
|\mathcal{O}_{13}\ra &=& (|t_1,1\ra + |t_2,2\ra+|t_3,3\ra)/\sqrt{3},\\
|\mathcal{O}_{14}\ra &=& (|t_1t_3t_2t_1,1\ra + |t_1t_3t_2t_1,2\ra+|t_1t_3t_2t_1,3\ra \\
&+& |t_2t_3t_1t_2,1\ra+ |t_2t_3t_1t_2,2\ra+|t_2t_3t_1t_2,3\ra)/\sqrt{6} .
\end{eqnarray*}
The unitary walk on the quotient graph can be written as
\be
\hat{U}_H=\hat{S}_H\cdot \hat{C}_H .
\ee
The matrices $\hat{S}_H$ and $\hat{C}_H$ are given by, $\hat{S}_H=\hat{S}'+\hat{S}'^\dag$ and $\hat{C}_H=1\oplus\hat{C}'\oplus\hat{C}''\oplus\hat{C}'\oplus\hat{C}''\oplus\hat{C}'\oplus 1$ where,
\begin{eqnarray*}
\hat{S}'&= & (|\mathcal{O}_1\ra\la\mathcal{O}_2| + |\mathcal{O}_3\ra\la\mathcal{O}_4| + |\mathcal{O}_5\ra\la\mathcal{O}_7| + |\mathcal{O}_6\ra\la\mathcal{O}_9| \\
&+& |\mathcal{O}_8\ra\la\mathcal{O}_{12}| + |\mathcal{O}_{10}\ra\la\mathcal{O}_{13}| + |\mathcal{O}_{11}\ra\la\mathcal{O}_{14}|) ,
\end{eqnarray*}
\[
\hat{C}'=\begin{pmatrix}-1/3 & 2\sqrt{2}/3\\ 2\sqrt{2}/3 & 1/3\end{pmatrix} 
\]
and
\[
\hat{C}''=\begin{pmatrix}-1/3 & 2/3 & 2/3 \\ 2/3 & -1/3 & 2/3 \\2/3 & 2/3 & -1/3\end{pmatrix} .
\]

\begin{figure}[h]
\begin{center}
\includegraphics[scale=0.4]{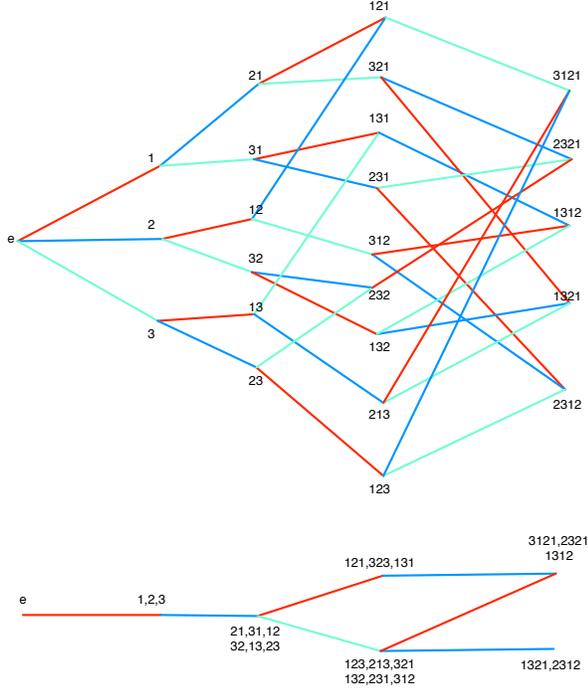}
\end{center}
\caption{(Color online) The graph $\Gamma(S_4,\{(1,2),(1,3),(1,4)\})$ and it quotient graph under the subgroup $H$.} \label{S_4_3gen}
\end{figure}

\end{example}

\begin{example}[Example 4.]
Consider the hypercube. The automorphism group of the hypercube is $\text{Aut}(\Gamma(\mathcal{Z}_2^n,Y))\simeq \mathcal{Z}_2^nS_n$. We focus on the subgroup $H_1=S_n$ and look at the resulting quotient graph. We consider the case when $n=3$, but the procedure for a general $n$ is very similar. The subgroup $H_1$ consists of all possible permutations of $n$ directions. The orbits under the action of this subgroup are given by,
\begin{eqnarray*}
|\mathcal{O}_1\ra &=& (|000,1\ra+|000,2\ra+|000,3\ra)/\sqrt{3}, \\
|\mathcal{O}_2\ra &=& (|001,1\ra+|010,2\ra+|100,3\ra)/\sqrt{3}, \\
|\mathcal{O}_3\ra &=& (|001\ra (|2\ra+|3\ra)+|010\ra(|1\ra+|3\ra)  \\
&+& |100\ra(|1\ra+|2\ra))/\sqrt{6}, \\
|\mathcal{O}_4\ra &=& (|011\ra(|2\ra+|3\ra)+|101\ra(|1\ra+|3\ra) \\
&+& |110\ra(|2\ra+|3\ra))/\sqrt{6}, \\
|\mathcal{O}_5\ra &=& (|011,3\ra+|101,2\ra+|110,1\ra)/\sqrt{3}, \\
|\mathcal{O}_6\ra &=& (|111,1\ra+|111,2\ra+|111,3\ra)/\sqrt{3} .
\end{eqnarray*}
The graph becomes a line as shown in Fig.~\ref{hypercube_line} and all the vertices of a certain Hamming weight collapse to a point. This fact had first been observed in \cite{MooRus02}. In \cite{SKW03},  this idea was used to construct a search algorithm on the hypercube. As observed  in \cite{SKW03}, the states on the line can be relabeled $|0,R\ra, |1,L\ra, |1,R\ra, |2,L\ra, |2,R\ra, |3,L\ra$. For the general hypercube of dimension $n$, these states generalize to
\begin{eqnarray}
|x,R\ra &=& \sqrt{\frac{1}{(n-x){n\choose x}}}\sum_{|\vec{x}|=x}\sum_{x_d=0} |\vec{x},d\ra, \nonumber \\
|x,L\ra &=& \sqrt{\frac{1}{(x){n \choose x}}}\sum_{|\vec{x}|=x}\sum_{x_d=1} |\vec{x},d\ra ,
\end{eqnarray}
where $|\vec{x}|$ is the Hamming weight of $\vec{x}$.

\begin{figure}[h]
\begin{center}
\includegraphics[scale=0.4]{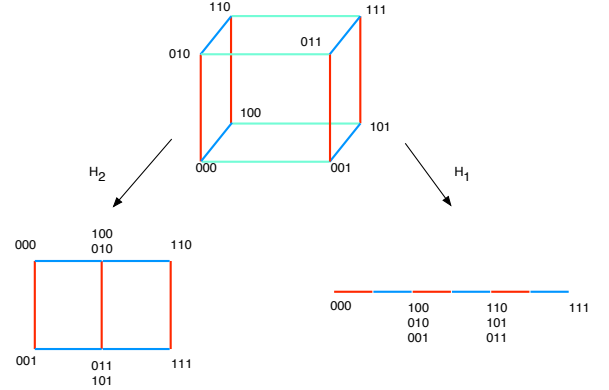}
\end{center}
\caption{(Color online) The $n=3$ hypercube and its quotient graphs.} \label{hypercube_line}
\end{figure}

Under the action of a different subgroup $H_2=S_{n-1}$ consisting of permutations of $n-1$ directions and the corresponding permutations of vertices, the quotient graph is shown in Fig. (\ref{hypercube_line}). The basis states of $\mathcal{H}_{H_2}$ when $n=3$, are
\begin{eqnarray*}
|\mathcal{O}_1\ra &=& (|000,1\ra, \\
|\mathcal{O}_2\ra & = & (|000,2\ra+|000,3\ra)/\sqrt{2}, \\
|\mathcal{O}_3\ra &=& (|001,1\ra, \\
|\mathcal{O}_4\ra &=& (|001,2\ra+|001,3\ra)/\sqrt{2}, \\
|\mathcal{O}_5\ra &=& (|010, 2\ra + |100, 3\ra)/\sqrt{2},  \\
|\mathcal{O}_6\ra &=& (|010,1\ra + |100, 1\ra)/\sqrt{2}, \\
|\mathcal{O}_7\ra &=& (|010,3\ra+|100,2\ra)/\sqrt{2}, \\
|\mathcal{O}_8\ra &=& (|011, 2\ra + |101, 3\ra)/\sqrt{2},  \\
|\mathcal{O}_9\ra &=& (|011,1\ra + |101, 1\ra)/\sqrt{2}, \\
|\mathcal{O}_{10}\ra &=& (|011,3\ra+|101,2\ra)/\sqrt{2}, \\
|\mathcal{O}_{11}\ra & = & (|110,2\ra+|110,3\ra)/\sqrt{2}, \\
|\mathcal{O}_{12}\ra &=& (|110,1\ra, \\
|\mathcal{O}_{13}\ra &=& (|111,2\ra+|111,3\ra)/\sqrt{2}, \\
|\mathcal{O}_{14}\ra &=& (|111,1\ra .
\end{eqnarray*}

\begin{figure}[h]
\begin{center}
\includegraphics[scale=0.4]{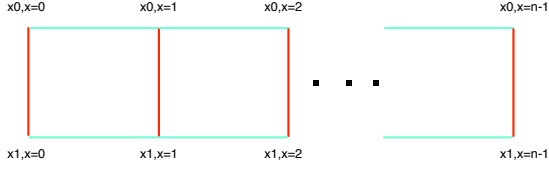}
\end{center}
\caption{(Color online) The quotient graph of a general hypercube under the group $S_{n-1}$.} \label{hypercube_planar}
\end{figure}

For any general $n$, the graph is still planar as shown in Fig.~\ref{hypercube_planar} and there will be $6n-4$ basis states. They can be labeled as $|x0, L\ra,|x0, R\ra, |x0,D\ra, |x1,L\ra,|x1,R\ra,|x1,U\ra$, where $x$ is the Hamming weight of the last $n-1$ bits (which fall under the action of the subgroup $S_{n-1}$) and the bit next to it is the first bit. $L,R,U$ and $D$ stand for left, right, up and down respectively. They are given by
\begin{eqnarray}
|x0,R\ra &=& \sqrt{\frac{1}{(n-1-x){n-1\choose x}}}\sum_{|\vec{x}|=x}\sum_{x_d=0} |\vec{x}0,d\ra, \nonumber \\
|x0,L\ra &=& \sqrt{\frac{1}{(x){n-1 \choose x}}}\sum_{|\vec{x}|=x}\sum_{x_d=1} |\vec{x}0,d\ra \nonumber \\
|x0,D\ra &=& \sqrt{\frac{1}{{n-1 \choose x}}}\sum_{|\vec{x}|=x} |\vec{x}0,1\ra \nonumber \\
|x1,R\ra &=& \sqrt{\frac{1}{(n-1-x){n-1\choose x}}}\sum_{|\vec{x}|=x}\sum_{x_d=0} |\vec{x}1,d\ra, \nonumber \\
|x1,L\ra &=& \sqrt{\frac{1}{(x){n-1 \choose x}}}\sum_{|\vec{x}|=x}\sum_{x_d=1} |\vec{x}1,d\ra \nonumber \\
|x1,U\ra &=& \sqrt{\frac{1}{{n-1 \choose x}}}\sum_{|\vec{x}|=x} |\vec{x}1,1\ra.
\end{eqnarray}
Note that the states $|x0,U\ra$ and $|x1,D\ra$ do not exist. Moreover,  $|x0,L\ra$ and $|x1,L\ra$ do not exist when $x=0$ and $|x0,R\ra$ and $|x1,R\ra$ do not exist when $x=n-1$. The unitary matrices describing the walk on these graphs are
\be
\hat{U}_{H_1}=\begin{pmatrix}
0 & -1/3 & 2\sqrt{2}/3 & 0 & 0 & 0 \\
1 & 0 & 0 & 0 & 0 & 0 \\
0 & 0 & 0 & -1/3 & 2\sqrt{2}/3 & 0 \\
0 & 2\sqrt{2}/3 & 1/3 & 0 & 0 & 0 \\
0 & 0 & 0 & 0 & 0 & 1 \\
0 & 0 & 0 & 2\sqrt{2}/3 & 1/3 & 0
\end{pmatrix} ,
\ee
and
\be
\hat{U}_{H_2}=\hat{S}_{H_2}\cdot \hat{C}_{H_2} .
\ee
The matrices $\hat{S}_{H_2}$ and $\hat{C}_{H_2}$ are given by, $\hat{S}_{H_2}=\hat{S}'+\hat{S}'^\dag$ and $\hat{C}_{H_2}=\hat{C}'\oplus \hat{C}'\oplus\hat{C}''\oplus\hat{C}''\oplus \hat{C}'\oplus\hat{C}'$ where,
\begin{eqnarray*}
\hat{S}'&= & (|\mathcal{O}_1\ra\la\mathcal{O}_3| + |\mathcal{O}_2\ra\la\mathcal{O}_5| + |\mathcal{O}_4\ra\la\mathcal{O}_8| + |\mathcal{O}_6\ra\la\mathcal{O}_9| \\
&+& |\mathcal{O}_7\ra\la\mathcal{O}_{11}| + |\mathcal{O}_{10}\ra\la\mathcal{O}_{13}| + |\mathcal{O}_{12}\ra\la\mathcal{O}_{14}|) ,
\end{eqnarray*}
\[
\hat{C}'=\begin{pmatrix}-1/3 & 2\sqrt{2}/3\\ 2\sqrt{2}/3 & 1/3\end{pmatrix} 
\]
and
\[
\hat{C}''=\begin{pmatrix}-1/3 & 2/3 & 2/3 \\ 2/3 & -1/3 & 2/3 \\2/3 & 2/3 & -1/3\end{pmatrix} .
\]
\end{example}

\begin{example}[Example 5.]
While we have shown how to construct quotient graphs for discrete-time walks on Cayley graphs, the idea of a quotient graph is more general.  In this example, we consider the ``glued trees" graph shown in Fig.~\ref{Glued_trees}. This graph is not regular and hence not a Cayley graph. It is undirected, and so we can easily define a continuous walk on it. Because the continuous walk does not have a coin space, we need not consider permutations of directions in the automorphisms. Quantum walks on this graph were first analyzed in \cite{CFG02}, and it has been shown that quantum walks move exponentially faster on this graph from ``entrance" to ``exit" than classical walks. The main reason for this exponential speed up is that the quantum walk moves in a superposition of all the vertices in a given column. It can be seen that in any given column, the vertices which branch out from the same vertex in the previous column can be interchanged as long as the corresponding interchange on the other side of the central column takes place. Therefore, the automorphism group of this graph is $Z_2^k$, where $k$ is one half of the total number of vertices on one side of the central column. Under the action of these automorphisms, the vertices in each column form a single orbit, and hence collapse to a single point in the quotient graph. There are $2n+1$ orbits under the action of this subgroup, where the columns $j$ are such that $0\leq j\leq 2n$. The orbits can be written as
\be
|\mathcal{O}_j\ra=2^{-\min[j,2n-j]/2} \sum_{v \in\ {\rm column}\ {j}} |v\ra .
\ee
The Hamiltonian for the quantum walk on the quotient graph becomes \cite{KLMW06}
\begin{eqnarray}\label{Hamiltonian_line}
  \langle \tilde{j}|H|\tilde{j} \pm 1\rangle &=& -\sqrt{2}\gamma \nonumber\\
 \langle \tilde{j}|H|\tilde{j}\rangle &=& \left\{ \begin{array}{ll}
    2\gamma & {j}=0,n,2n \\
    3\gamma & {\rm otherwise,} \\
    \end{array} \right.
\end{eqnarray}
with all other matrix elements zero.  This is also shown in Fig.~\ref{Glued_trees} where the $\gamma$ has been dropped for brevity.

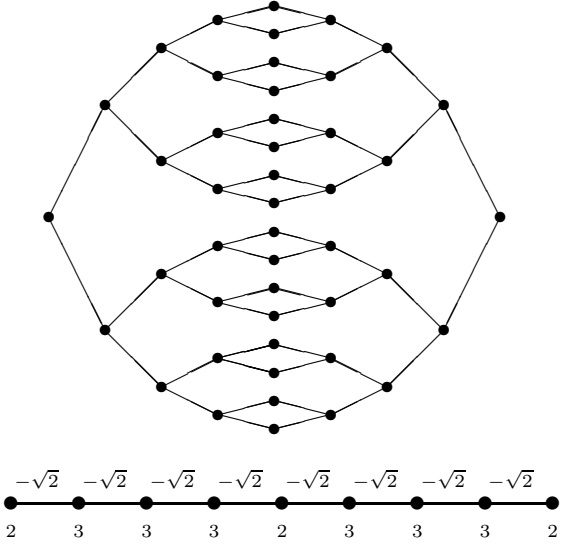
\begin{figure}[ht]
    \centering
    \setlength{\unitlength}{0.75cm}
    \begin{picture}(4,7.5)
        \put(-2,3.75){\line(1,2){1}}
        \put(-2,3.75){\line(1,-2){1}}
        \put(-1,1.75){\line(1,-1){1}}
        \put(-1,1.75){\line(1,1){1}}
        \put(-1,5.75){\line(1,1){1}}
        \put(-1,5.75){\line(1,-1){1}}
        \put(0,6.75){\line(2,1){1}}
        \put(0,6.75){\line(2,-1){1}}
        \put(0,4.75){\line(2,1){1}}
        \put(0,4.75){\line(2,-1){1}}
        \put(0,0.75){\line(2,1){1}}
        \put(0,0.75){\line(2,-1){1}}
        \put(0,2.75){\line(2,1){1}}
        \put(0,2.75){\line(2,-1){1}}
        \put(1,7.25){\line(4,1){1}}
        \put(1,7.25){\line(4,-1){1}}
        \put(1,6.25){\line(4,1){1}}
        \put(1,6.25){\line(4,-1){1}}
        \put(1,5.25){\line(4,1){1}}
        \put(1,5.25){\line(4,-1){1}}
        \put(1,4.25){\line(4,1){1}}
        \put(1,4.25){\line(4,-1){1}}
        \put(1,1.25){\line(4,1){1}}
        \put(1,1.25){\line(4,-1){1}}
        \put(1,0.25){\line(4,1){1}}
        \put(1,0.25){\line(4,-1){1}}
        \put(1,1.25){\line(4,1){1}}
        \put(1,1.25){\line(4,-1){1}}
        \put(1,2.25){\line(4,1){1}}
        \put(1,2.25){\line(4,-1){1}}
        \put(1,3.25){\line(4,1){1}}
        \put(1,3.25){\line(4,-1){1}}
        \put(6,3.75){\line(-1,2){1}}
        \put(6,3.75){\line(-1,-2){1}}
        \put(5,1.75){\line(-1,-1){1}}
        \put(5,5.75){\line(-1,1){1}}
        \put(5,1.75){\line(-1,1){1}}
        \put(5,5.75){\line(-1,-1){1}}
        \put(4,6.75){\line(-2,1){1}}
        \put(4,6.75){\line(-2,-1){1}}
        \put(4,4.75){\line(-2,1){1}}
        \put(4,4.75){\line(-2,-1){1}}
        \put(4,0.75){\line(-2,1){1}}
        \put(4,0.75){\line(-2,-1){1}}
        \put(4,2.75){\line(-2,1){1}}
        \put(4,2.75){\line(-2,-1){1}}
        \put(3,7.25){\line(-4,1){1}}
        \put(3,7.25){\line(-4,-1){1}}
        \put(3,6.25){\line(-4,1){1}}
        \put(3,6.25){\line(-4,-1){1}}
        \put(3,5.25){\line(-4,1){1}}
        \put(3,5.25){\line(-4,-1){1}}
        \put(3,4.25){\line(-4,1){1}}
        \put(3,4.25){\line(-4,-1){1}}
        \put(3,0.25){\line(-4,1){1}}
        \put(3,0.25){\line(-4,-1){1}}
        \put(3,1.25){\line(-4,1){1}}
        \put(3,1.25){\line(-4,-1){1}}
        \put(3,2.25){\line(-4,1){1}}
        \put(3,2.25){\line(-4,-1){1}}
        \put(3,3.25){\line(-4,1){1}}
        \put(3,3.25){\line(-4,-1){1}}
        \put(6,3.75){\circle*{0.2}}
        \put(-2,3.75){\circle*{0.2}}
        \put(5,5.75){\circle*{0.2}}
        \put(-1,5.75){\circle*{0.2}}
        \put(5,1.75){\circle*{0.2}}
        \put(-1,1.75){\circle*{0.2}}
        \put(4,6.75){\circle*{0.2}}
        \put(4,0.75){\circle*{0.2}}
        \put(0,6.75){\circle*{0.2}}
        \put(0,0.75){\circle*{0.2}}
        \put(4,4.75){\circle*{0.2}}
        \put(0,4.75){\circle*{0.2}}
        \put(4,2.75){\circle*{0.2}}
        \put(0,2.75){\circle*{0.2}}
        \put(3,7.25){\circle*{0.2}}
        \put(1,7.25){\circle*{0.2}}
        \put(3,0.25){\circle*{0.2}}
        \put(1,0.25){\circle*{0.2}}
        \put(3,6.25){\circle*{0.2}}
        \put(3,1.25){\circle*{0.2}}
        \put(1,6.25){\circle*{0.2}}
        \put(1,1.25){\circle*{0.2}}
        \put(3,5.25){\circle*{0.2}}
        \put(3,2.25){\circle*{0.2}}
        \put(1,5.25){\circle*{0.2}}
        \put(1,2.25){\circle*{0.2}}
        \put(3,4.25){\circle*{0.2}}
        \put(3,3.25){\circle*{0.2}}
        \put(1,4.25){\circle*{0.2}}
        \put(1,3.25){\circle*{0.2}}
        \put(2,7.5){\circle*{0.2}}
        \put(2,7){\circle*{0.2}}
        \put(2,6.5){\circle*{0.2}}
        \put(2,6){\circle*{0.2}}
        \put(2,5.5){\circle*{0.2}}
        \put(2,5){\circle*{0.2}}
        \put(2,4.5){\circle*{0.2}}
        \put(2,4){\circle*{0.2}}
        \put(2,3.5){\circle*{0.2}}
        \put(2,3){\circle*{0.2}}
        \put(2,2.5){\circle*{0.2}}
        \put(2,2){\circle*{0.2}}
        \put(2,1.5){\circle*{0.2}}
        \put(2,1){\circle*{0.2}}
        \put(2,0.5){\circle*{0.2}}
        \put(2,0){\circle*{0.2}}
    \end{picture}
\vspace{.2in}
    \setlength{\unitlength}{0.9cm}
    \begin{picture}(9,1)
        \put(0.55,0.5){\line(1,0){8}}
        \put(0.55,0.5){\circle*{0.2}}
        \put(1.55,0.5){\circle*{0.2}}
        \put(2.55,0.5){\circle*{0.2}}
        \put(3.55,0.5){\circle*{0.2}}
        \put(4.55,0.5){\circle*{0.2}}
        \put(5.55,0.5){\circle*{0.2}}
        \put(6.55,0.5){\circle*{0.2}}
        \put(7.55,0.5){\circle*{0.2}}
        \put(8.55,0.5){\circle*{0.2}}
        \put(0.6,0.75){{\scriptsize$-\sqrt{2}$}}
        \put(1.6,0.75){{\scriptsize$-\sqrt{2}$}}
        \put(2.6,0.75){{\scriptsize$-\sqrt{2}$}}
        \put(3.6,0.75){{\scriptsize$-\sqrt{2}$}}
        \put(4.6,0.75){{\scriptsize$-\sqrt{2}$}}
        \put(5.6,0.75){{\scriptsize$-\sqrt{2}$}}
        \put(6.6,0.75){{\scriptsize$-\sqrt{2}$}}
        \put(7.6,0.75){{\scriptsize$-\sqrt{2}$}}
        \put(0.475,0){{\scriptsize$2$}}
        \put(1.475,0){{\scriptsize$3$}}
        \put(2.475,0){{\scriptsize$3$}}
        \put(3.475,0){{\scriptsize$3$}}
        \put(4.475,0){{\scriptsize$2$}}
        \put(5.475,0){{\scriptsize$3$}}
        \put(6.475,0){{\scriptsize$3$}}
        \put(7.475,0){{\scriptsize$3$}}
        \put(8.475,0){{\scriptsize$2$}}
    \end{picture}
\caption{The glued trees graph and its quotient graph.}
\label{Glued_trees}
\end{figure}

\end{example}


\subsection{Automorphism group of the quotient graph}
In this section we determine the automorphisms of the quotient graph which are induced from the automorphism group of the original graph and the subgroup used to obtain the quotient graph. In dealing with the automorphisms of $\Gamma$ we used permutations of vertices and edges, and this in turn corresponds to permutations of basis vectors which preserves the shift matrix. On the quotient graph, we define those permutations of {\it orbits} which preserve the $\hat{S}_H$ matrix as automorphisms, since there is no natural choice of edge colors. These permutations of orbits which preserve the new shift matrix also preserve the quotient graph.

Let $G_1$ be a set of automorphisms of $\Gamma$ which are of the following type. If they take a basis vector belonging to a $H$-orbit $\mathcal{O}_1$ to a basis vector belonging to $\mathcal{O}_2$, then they take every basis vector in $\mathcal{O}_1$ to some basis vector in $\mathcal{O}_2$. Clearly, all the automorphisms in $H$ are of this type, under the special case when $\mathcal{O}_1=\mathcal{O}_2$. It is also easy to verify that $G_1$ is a subgroup of $G$, and that $H$ is a subgroup of $G_1$. We now show that $H$ is a normal subgroup of $G_1$ i.e., $g h g^{-1} \in H$, $\forall g\in G_1$ and $\forall h\in H$.

\begin{theorem}
Given the group $G_1$ defined as above we have,
\begin{enumerate}
\item The subgroup $H$ is a normal subgroup of $G_1$.
\item $G_1$ is the largest subgroup of $G$ such that $H$ is a normal subgroup of $G_1$---that is, for any $g\in G$, if $g h g^{-1} \in H$ $\forall h\in H$ then $g\in G_1$.
\end{enumerate}
\end{theorem}
\begin{proof}
We show this by considering the action of all of these group elements on the set of basis vectors. 
\begin{enumerate}
\item Let $x$ be any basis element belonging to some $H$-orbit $\mathcal{O}_1$, and let $g$ take every element in $\mathcal{O}_2$ to some element in $\mathcal{O}_1$. Then $g h g^{-1} x = g h y$, where $y\in \mathcal{O}_2$. Now, $h y = z$ where $z\in \mathcal{O}_2$ since these orbits are formed under the action of $H$. Hence, $g z = x'$, where $x'\in \mathcal{O}_1$. But every $x'\in \mathcal{O}_1$ can be written as $h' x$ for some $h'\in H$. Thus, $g h g^{-1} \in H$. 
\item Consider some basis element $x \in \mathcal{O}_3$ and let $g x \in \mathcal{O}_4$. Since $g h x = h' g x$, $g (h x) = h' y = y'$, where $y,y' \in \mathcal{O}_4$. Therefore, $g$ takes $h x \in \mathcal{O}_3$ to $y'\in \mathcal{O}_4$, but since $h\in H$ is arbitrary, $g$ takes every element of $\mathcal{O}_3$ to some element of $\mathcal{O}_4$. It follows that $g\in G_1$.

\end{enumerate}
\end{proof}
Since $H$ is normal in $G_1$, the quotient set $G_1/H$, i.e., the set of all cosets $g H =\{g h | g\in H\}$, is a group. This group has a natural representation in the Hilbert space $\mathcal{H}_H$ as a permutation matrix in the basis where each orbit is a basis vector. 
\begin{theorem}
$G_1/H \subset \text{Aut}(\Gamma_H)$. 
\end{theorem}
\begin{proof}
Consider any automorphism $g\in G_1$ and let $\sigma(g)$ be its representation in $\mathcal{H}$. Then, we have $\sigma(g) S \sigma(g^{-1}) = S$. The projection of this into $\mathcal{H}_H$ is given by 
\begin{equation}
P_H \sigma(g) S \sigma(g^{-1}) P_H = P_H S P_H=S_H .
\end{equation}
The representation $\sigma(g)$ commutes with $P_H$, since it permutes all the vectors in an orbit to vectors in another orbit. Therefore,
\begin{equation}
P_H \sigma(g)P_H SP_H \sigma(g^{-1}) P_H =S_H .
\end{equation}
But as a representation, $P_H \sigma(g) P_H = \sigma(gH)$. This means that the representation of $gH \in G_1/H$ in $\mathcal{H}_H$ is a group of symmetries of $S_H$ and therefore $G_1/H \subset \text{Aut}(\Gamma)$.
\end{proof}
We see that the quotient graph is obtained from $\Gamma$ modulo the symmetries in $H$.

\section{Hitting time}
\subsection{Definition}
The hitting time $\tau_h$ of a classical random walk is defined as the average time for the walk to hit a designated `final' vertex $v_f$ given that the walk began with some initial distribution $p_i$:
\begin{equation}\label{ht1}
\tau_h = \sum_{t=0}^\infty t p(t),
\end{equation}
where $p(t)$ is the probability of being in the final vertex for the first time at time step $t$. In order to carry this notion of hitting time over to the quantum case, we need to make the meaning of $p(t)$ more precise. In particular, we need to define clearly what ``for the first time'' means for a quantum walk.  As described in \cite{KB05}, we do this by performing a measurement of the particle at every step of the walk to see if the particle has reached the final vertex or not. The measurement $M$ which is used has projectors $\hat{P}_f$ and $\hat{Q}_f=\hat{I}-\hat{P}_f$ representing the particle being found or not found at the final vertex, respectively.  The projector is defined $\hat{P}_f=|x_f\ra\la x_f|\otimes\hat{I}_c$, where $|x_f\ra$ is the final vertex state and $\hat{I}_c$ is the identity operator on the coin space. Using this definition, each step of the measured walk consists of an application of the unitary evolution operator $\hat{U}$ followed by the measurement $M$.

By including these measurements at each step we can use the same expression (\ref{ht1}) for the hitting time of the quantum walk, where the probability $p(t)$ becomes
\begin{equation}\label{prob.eqn}
p(t)=\tr\{\hat{P}_f\hat{U}[\hat{Q}_f\hat{U}]^{t-1}
\rho_0[\hat{U^{\dag}}\hat{Q}_f]^{t-1}\hat{U^{\dag}}\hat{P}_f\} .
\end{equation}
To sum the series (\ref{ht1}) explicitly using the expression for $p(t)$ in Eq.~(\ref{prob.eqn}), we rewrite the expression in terms of {\it superoperators} (linear transformations on operators) $\mathcal{N}$ and $\mathcal{Y}$, defined by
\begin{eqnarray}\label{superops}
\mathcal{N}\rho = \hat{Q}_f\hat{U}\rho\hat{U^{\dag}}\hat{Q}_f \nonumber\\
\mathcal{Y}\rho=\hat{P}_f\hat{U}\rho\hat{U^{\dag}}\hat{P}_f.
\end{eqnarray}
In terms of $\mathcal{N}$ and $\mathcal{Y}$,
$p(t)=\tr\{\mathcal{Y}\mathcal{N}^{t-1}\rho_0\}$.
We introduce a new superoperator $\mathcal{O}(l)$ which depends on a real parameter $l$:
\begin{equation}
\mathcal{O}(l) = l\sum_{t=1}^\infty (l\mathcal{N})^{t-1} ,
\label{superop_sum}
\end{equation}
which is a function of a parameter $l$.  The hitting time now becomes
\begin{equation}
\tau_h = \frac{d}{dl} \tr\{ \mathcal{Y} \mathcal{O}(l) \rho_0 \} \biggr|_{l=1}.
\label{derivative_form}
\end{equation}

If the superoperator $\mathcal{I}-l\mathcal{N}$ is invertible, then we can replace the sum (\ref{superop_sum}) with the closed form
\begin{equation}
\mathcal{O}(l) = l(\mathcal{I}-l\mathcal{N})^{-1}.
\end{equation}
The derivative in (\ref{derivative_form}) is
\begin{equation}
\frac{d\mathcal{O}}{dt}(1)
= (\mathcal{I}-\mathcal{N})^{-1}+\mathcal{N}(\mathcal{I}-\mathcal{N})^{-2}
= (\mathcal{I}-\mathcal{N})^{-2}.
\end{equation}
This gives us the following expression for the hitting time:
\begin{equation}
\tau_h = \tr\{\mathcal{Y}(\mathcal{I}-\mathcal{N})^{-2}\rho_0\}.
\label{closed_form_tau}
\end{equation}

To evaluate (\ref{closed_form_tau}), we write these superoperators as matrices using Roth's lemma \cite{Roth34}. As shown in \cite{KB05}, we can then {\it vectorize} the density operators and operators on states, and write the action of superoperators as simple matrix multiplication.  Any matrix can be vectorized by turning its rows into columns and
stacking them up one by one, so that a $D\times D$ matrix becomes
a column vector of size $D^2$. Consequently the superoperators become matrices
of size $D^2\times D^2$. This method of vectorization takes operators on one Hilbert space $\mathcal{H}$ to vectors in another Hilbert space $\mathcal{H}'=\mathcal{H}\otimes\mathcal{H}^\ast$ and so superoperators in $\mathcal{H}$ are operators in $\mathcal{H}'$. Note that a basis $\{|u_{i j}\ra\}$ for $\mathcal{H}'$ can be obtained from a basis $\{|v_i\ra\}$ for $\mathcal{H}$ by defining
\begin{equation}
|u_{i j}\ra=|v_i\ra\otimes |v_j\ra^\ast .
\end{equation}
For our superoperators $\mathcal{N}$ and $\mathcal{Y}$ we then get
\begin{eqnarray}
(\mathcal{N}\rho)^v &=& \left[ (\hat{Q}_f\hat{U})\otimes(\hat{Q}_f\hat{U})^\ast \right] \rho^v ,
\nonumber \\
(\mathcal{Y}\rho)^v &=& \left[ (\hat{P}_f\hat{U})\otimes(\hat{P}_f\hat{U})^\ast \right] \rho^v .
\end{eqnarray}
Let $\mathbf{N}=(\hat{Q}_f\hat{U})\otimes(\hat{Q}_f\hat{U})^\ast$ and $\mathbf{Y}=(\hat{P}_f\hat{U})\otimes(\hat{P}_f\hat{U})^\ast$. The hitting time becomes
\begin{equation}\label{hittime}
\tau_h = I^v \cdot \left( \mathbf{Y}(\mathbf{I}-\mathbf{N})^{-2}\rho^v \right) .
\end{equation}
Using this vectorization transformation, we treat the superoperators as operators on a larger Hilbert space and thus can find their inverses. However, the expression in Eq.~(\ref{hittime}) is not always well defined, because the matrix $\mathbf{I}-\mathbf{N}$ may not be invertible. In \cite{KB06}, it is shown that when this matrix is not invertible, then the quantum walk has an infinite hitting time for certain initial states. An infinite hitting time means that the probability that the particle reaches the final vertex at any time step (i.e., $\sum_{t=0}^\infty p(t)$) is less than unity. The projector $\hat{P}$ onto all initial states that never reach the final vertex is non-zero whenever the matrix $\mathbf{I}-\mathbf{N}$ is non-invertible. The relation between the null space of $\mathbf{I}-\mathbf{N}$ and the projector $\hat{P}$ is somewhat subtle, and is discussed in \cite{KB06}. Here, we show that if such a projector exists then it will give infinite hitting times. We begin by forming the projector $\hat{P}$ onto the subspace spanned by all eigenstates of $\hat{U}$ which have no overlap with the final vertex.  This projector is orthogonal to the projector onto the final vertex, $\hat{P}\hat{P}_f = \hat{P}_f\hat{P} = 0$, and commutes with $\hat{U}$, $[\hat{U},\hat{P}]=0$. We can write any initial state as a superposition of a state in the subspace projected onto by $\hat{P}$ and a state orthogonal to it, giving the decomposition
\begin{equation}
|\Psi\ra=\hat{P} |\Psi\ra + (\hat{I}-\hat{P}) |\Psi\ra .
\end{equation}
It is easy to see that if $|\Psi\ra$ lies entirely inside $\hat{P}$, i.e., $\hat{P}|\Psi\ra=|\Psi\ra$, then under the unitary evolution the subsequent states will never have any component in the final vertex, and the probability defined in Eq.~(\ref{prob.eqn}) will be zero. Indeed, since $[\hat{P},\hat{U}]=0$ and $[\hat{P},\hat{Q}_f]=0$,
\begin{eqnarray*}
p(t) &=& \tr\{\hat{P}_f\hat{U}[\hat{Q}_f\hat{U}]^{t-1}
\rho_0[\hat{U^{\dag}}\hat{Q}_f]^{t-1}\hat{U^{\dag}}\hat{P}_f\}\\
&=& \tr\{\hat{P}_f\hat{U}[\hat{Q}_f\hat{U}]^{t-1}
\hat{P}\rho_0\hat{P}[\hat{U^{\dag}}\hat{Q}_f]^{t-1}\hat{U^{\dag}}\hat{P}_f\}\\
&=& \tr\{\hat{P}_f\hat{P}\hat{U}[\hat{Q}_f\hat{U}]^{t-1}
\rho_0[\hat{U^{\dag}}\hat{Q}_f]^{t-1}\hat{U^{\dag}}\hat{P}^{\dag}\hat{P}_f\}\\
&=& 0 ,
\end{eqnarray*}
where $\rho_0=|\Psi\ra\la\Psi|$. Therefore, the hitting time for this initial state is infinite. More generally, if $|\Psi\ra$ has nonzero overlap with $\hat{P}$, $\hat{P}|\Psi\ra \ne 0$, then that component of $|\Psi\ra$ can never reach the final vertex. The probability of ever hitting the final vertex if one starts with this initial state is
\begin{equation}
p=|\la\Psi|(\hat{I}-\hat{P})|\Psi\ra|^2 < 1,
\end{equation}
and the hitting time is again infinite.

To construct this projector, we look at the spectral decomposition of $\hat{U}$. If $\hat{U}$ has at least one sufficiently degenerate eigenspace, then we can construct a subspace of this eigenspace which has a zero overlap with the final vertex. For instance, consider one such degenerate eigenspace which has a degeneracy of $k$. Since the vector space at the final vertex is $d$ dimensional (i.e., it has $d$ coin degrees of freedom), we would be solving the following $d\times k$ system of homogeneous equations:
\begin{equation}
a_1\begin{pmatrix}
v_{N-d+1}^1\\
\vdots\\
v_N^1\end{pmatrix}+
a_2\begin{pmatrix}
v_{N-d+1}^2\\
\vdots\\
v_N^2\end{pmatrix}+
\cdots+
a_k\begin{pmatrix}
v_{N-d+1}^k\\
\vdots\\
v_N^k\end{pmatrix}=0 .
\end{equation}
Here we use a labeling where the final vertex in some coin state occupies the last $d$ entries of the eigenvectors. The subscript refers to the component of the eigenvector, and the superscript distinguishes the eigenvectors in the degenerate eigenspace. This system is under-determined if $k>d$, and it will always have a nontrivial solution---in fact, it will have a space of solutions of dimension $k-d$. Therefore, it is sufficient that there exist at least one eigenspace of $\hat{U}$ with dimension greater than the dimension of the coin, in order to have a nonzero projector $\hat{P}$.  If there is more than one degenerate eigenvalue with multiplicity greater than $d$, the subspace projected onto by $\hat{P}$ will include {\it all} the eigenvectors of $\hat{U}$ which have no overlap with the final vertex. This means that the size of this projector will be $\sum_i (d_i-d)$ where the sum runs over all the eigenspaces of $\hat{U}$ which have a degeneracy $d_i> d$. Here we assume that $\hat{U}$ has no eigenvectors in the other eigenspaces which happen by chance to have no overlap with the final vertex, even though the eigenspace is not sufficiently degenerate to construct such an eigenstate.  Such ``accidentally infinite'' hitting times, like accidental degeneracies, are presumably rare. The phenomenon of infinite hitting times for quantum walks has no classical analogue since classical random walks always reach the final vertex eventually if the graph is connected.

Finally, we answer the question: given any vertex $v$ on the graph, it is natural to ask if there exists any superposition of its coin states which overlaps with $\hat{P}$, for which coin state the overlap is maximum, and for which it is minimum (or zero). We write the projector $\hat{P}$ in the form
\begin{equation}
\hat{P}=\sum_{i,j,k,l} A_{i,j,k,l}|x_i\ra\la x_j|\otimes |k\ra\la l|,
\end{equation}
where $\{|x_i\ra\}$ are the vertices and $\{|k\ra\}$ are the directions. Suppose the initial state is
\begin{equation}\label{alpha}
|\Psi\ra=|v\ra\otimes\sum_i\alpha_i|i\ra = |v\ra \otimes |\alpha\ra.
\end{equation}
Its overlap with the projector $\hat{P}$ is given by,
\begin{equation}
\la\Psi|\hat{P}|\Psi\ra=\sum_{k,l} A_{v,v,k,l} \alpha_k^*\alpha_l .
\end{equation}
To find the superposition of coin states such that the overall initial state has the least (or greatest) overlap with $\hat{P}$, define the matrix,
\begin{equation}\label{C_matrix}
\hat{C}_v=\tr_{\rm vertices}\{\hat{P}|v\ra\la v|\otimes\hat{I}_{coin}\},(\hat{C}_v)_{kl}=A_{v,v,k,l} .
\end{equation}
The overlap of the initial state with $\hat{P}$ can be written in terms of this matrix as,
\begin{equation}
\la\Psi|\hat{P}|\Psi\ra=\la\alpha|\hat{C}_v|\alpha\ra .
\end{equation}
So now assuming that $\{\lambda_i,|e_i\ra\}$ is the spectral decomposition of $\hat{C}_v$, we can rewrite the overlap as,
\begin{equation}
\la\Psi|\hat{P}|\Psi\ra=\sum_i \lambda_i |\la\alpha|e_i\ra|^2 .
\end{equation}
The matrix $\hat{C}$ is Hermitian and positive, and hence has a spectral decomposition into a complete orthonormal basis of eigenvectors with non-negative eigenvalues. Assuming that $\{\lambda_i, e_i\}$ is the spectral decomposition of $\hat{C}_v$, we can write the overlap as 
\be
\la\Psi|\hat{P}|\Psi\ra=\sum_i \lambda_i |\la \alpha|e_i\ra|^2
\ee
From the above expression, we see that the overlap is maximum (or minimum) if $|\alpha\ra$ is in the direction of the eigenvector with the largest (or smallest) eigenvalue and zero if $|\alpha\ra$ is along the eigenvector with a zero eigenvalue. Therefore, if $\hat{C}_v$ does not have a zero eigenvalue (i.e., is positive definite), then for that vertex every superposition of coin states will overlap with $\hat{P}$. In other words, the hitting time will be infinite if one starts at that vertex no matter what coin state one chooses.

\subsection{Hitting time on quotient graphs}
In this subsection, we address the question of when quantum walks on quotient graphs have infinite hitting times. It is possible that for some subgroups, the walk on the quotient graph does not have infinite hitting times even if the walk on the original graph does. In order to carry over the discussion of hitting times to quotient graphs, we must keep in mind that the evolution operator is now followed by a measurement. To remain on the quotient graph (i.e., in the subspace given by $\hat{P}_H$), the measurement operators must commute with the symmetry operators $\sigma(h)$, $h \in H$.

If this condition is satisfied, then we can obtain a condition to check whether the quotient graph has initial states with infinite hitting times:  if the subspace of those initial states with infinite hitting times on the original graph whose projector is $\hat{P}$, has no nontrivial intersection with the subspace whose projector is $\hat{P}_H$ i.e., $\hat{P} \cap \hat{P}_H = \varnothing$, then the walk on the quotient graph does not have infinite hitting times.  If there is a nontrivial intersection, then it does.  (Here we have used the projectors onto the subspaces to denote the spaces themselves.) This condition can also be verified by obtaining the restriction of the evolution operator and the measurement operators onto the quotient graph. By diagonalizing the new unitary evolution operator and constructing the projector of states $\tilde{P}$ which have no overlap with the new final vertex state. The subspace of these states is exactly the intersection $\hat{P}\cap \hat{P}_H$. We will examine this condition for some of the examples considered above.

In the first example, we choose the final vertex to be $t_1t_2t_1$. The measurement operators are $\hat{P}_f=|t_1t_2t_1\ra\la t_1t_2t_1|\otimes \hat{I}$ and $\hat{I}-\hat{P}_f$. This measurement commutes with the subgroup chosen, and the quotient graph does not have infinite hitting times. This is because the original graph does not have infinite hitting times either i.e., $\hat{P}=\varnothing$.

For the second example, for the graph $\Gamma(S_3,\{(1,2),(2,3),(1,3)\})$, we used three different subgroups and form their quotient graphs. In order to determine whether the quotient graph has infinite hitting times for various subgroups, we must choose different final vertices for the different subgroups since the measurement must commute with the symmetries. Therefore, for $H_1$ we choose $t_1t_2$ as the final vertex, and the measurement operators are $|t_1t_2\ra\la t_1t_2|\otimes\hat{I}$ and its orthogonal complement. This measurement commutes with the subgroup $H_1$. For this final vertex and measurement, the original graph has infinite hitting times i.e., $\hat{P} \neq \varnothing$ and the quotient graph also has infinite hitting times i.e., $\hat{P}\cap \hat{P}_{H_1} \neq \varnothing$. In fact, using the $C$-matrix defined above in Eq.~(\ref{C_matrix}), we find that if the initial vertex is the identity $|e\ra$, then there is no superposition of coin states that has a finite hitting time, because $\hat{C}_v$ does not have a zero eigenvalue for $v=e$.

For the subgroup $H_2$, we choose the final vertices to be $t_1t_2$ and $t_2t_1$. Therefore, the measurement on the original graph must be a projective measurement with outcomes $\hat{P}_f=(|(t_1t_2\ra \la t_1t_2|+|t_2t_1\ra\la t_2t_1|)\otimes \hat{I}$ and its orthogonal complement. For this measurement and final vertices, the original graph has $\hat{P} = \varnothing$. Therefore, the quotient graph also does not have infinite hitting times. For the subgroup $H_3$, the measurement operators are $\hat{P}_f=(|(t_1t_2\ra \la t_1t_2|+|t_2t_1\ra\la t_2t_1|)\otimes \hat{I}$ and its orthogonal complement. For this measurement, neither the original graph nor the quotient graph have infinite hitting times.

In example 3, for the Cayley graph $\Gamma(S_4,\{(1,2),(1,3),(1,4)\})$, we choose the final vertices to be $|t_1t_3t_2t_1\ra$ and $|t_2t_3t_1t_2\ra$.  In this case, we find that while the original graph has infinite hitting times, the quotient graph does not. In fact, on the original graph, the equal superposition of all coin states at the vertex $|e\ra$ is the {\it only} superposition which does not have an infinite hitting time (i.e., the $\hat{C}_v$ matrix has only one zero eigenvalue with the equal superposition of coin states as its eigenvector).  It is precisely this vector which is included in the subspace of the quotient graph.  This is not a coincidence---in both cases, it is picked out by the symmetries of the graph.

In example 4, for the two different subgroups of the automorphism group considered for the hypercube, we find that the behavior of hitting times is very different. For the subgroup $H_1$, the quotient graph becomes a line with the vertex $\vec{0}=00\dots 0$ on one end and the vertex $\vec{1}=11\dots 1$ on the other. If one designates the final vertex to be $\vec{1}$ by choosing the measurement operators to be $\hat{P}_f=|\vec{1}\ra\la\vec{1}|\otimes\hat{I}$ and its orthogonal complement, then we find that this quotient graph does not have infinite hitting times for any initial state:  $\hat{P}\cap\hat{P}_{H_1}=\varnothing$. In fact, if the initial state is $|00\dots 0\ra\otimes\frac{1}{d}\sum_i |i\ra$, then the hitting time is polynomial in $d$, the dimension of the hypercube \cite{Kem03b,KB05}. Using the $C$-matrix for the original graph, we find that if the initial vertex is $00\dots 0$, then the equal superposition of all directions is the {\it only} zero eigenvector of $C_v$, which means that it is the only coin state that does {\it not} have an infinite hitting time.

On the other hand, choosing $11\dots 10$ as the final vertex and using the subgroup $H_2$, we find that the quotient graph shown in Fig.~\ref{hypercube_line} does have infinite hitting times for some initial states i.e., $\hat{P}\cap\hat{P}_{H_2}\neq\varnothing$. Using the $C$-matrix again, we find that if the initial vertex is $00\dots 0$, then the equal superposition of all directions once again is the only coin state that has no infinite hitting times. For every other superposition of coin states for that vertex (i.e., every other $|\alpha\ra$ in Eq.~(\ref{alpha})) $C_v$ has a nonzero eigenvalue.

\section{Discussion}

We have investigated the behavior of quantum walks on undirected graphs by making use of the automorphism group of the graph. Automorphisms of the graph may become symmetries of the discrete quantum walk, depending on the symmetries of the coin matrix. Quantum walks which respect the symmetries of some subgroup $H$ of this automorphism group have an invariant subspace in the total Hilbert space. We showed that the walk restricted to this subspace can be seen as a (different) quantum walk on a {\it quotient} graph, and that this graph can be constructed from the original graph given the subgroup $H$. The dynamics of the new walk can also be derived from the original walk and the subgroup. The quotient graph is obtained from the original graph by identifying vertices and edges which form an orbit under the action of $H$; this means that the quotient graph and the new quantum walk both have no symmetries coming from $H$. The new quantum walk only has the remaining automorphisms as its possible symmetries, and so it has, in a sense, ``used up'' the ones in $H$.

To discuss hitting times, we use the measured walk defined in \cite{Kem03b} and \cite{KB05}, which consists of the application of a unitary operator followed by a projective measurement at each time step. For the walk on the quotient graph to be preserved, the choice of measurement must commute with the symmetries in $H$.  This restriction is very important; even if the walk and initial state both have a larger group of symmetries, the walk will be on a quotient graph corresponding to a smaller subgroup $H$ if the measurement does not commute with the remaining elements of the larger group.

For instance, in Example 2, using the subgroup $H_1$, we obtained a walk on its quotient graph. Suppose the measurement is a projective measurement of the vertex $t_1t_2$. The initial state $|e\ra\otimes (|1\ra+|2\ra+|3\ra)/\sqrt(3)$, and the walk with the Grover coin $\hat{U}$ both have all the symmetries of  $H_2$.  But the measurement, which commutes with all the elements of $H_1$, does {\it not} commute with all those in $H_2$, and the effective walk will be on the quotient graph corresponding to $H_1$.

The remaining symmetries of the evolution operator can lead to degeneracy in its eigenspectrum \cite{KB06}, and may result in infinite hitting times on the quotient graph. In general, we found a condition to determine whether the walk on the quotient graph will have infinite hitting times:  given the original graph, the quantum walk and any subgroup $H$, one can determine the projector onto states with infinite hitting time $\hat{P}$, and the invariant subspace of the quotient graph $\hat{P}_H$.  If $\hat{P} \cap \hat{P}_H = \varnothing$ then the quotient graph does not have infinite hitting times.

Even when the hitting time is not infinite for an initial state on the quotient graph, it is possible that it could be extremely long.  It would be useful to have a criterion to pick out subgroups of the automorphism group whose quotient graphs have exponentially fast hitting times. For example, in the case of the hypercube, using the subgroup  $H_1$ (whose quotient graph is a line) turns out to give very fast hitting times. But on a general undirected graph it is not easy to determine whether there is a subgroup whose quotient graph gives fast hitting times.

To investigate this, we need to make the notion of ``fast" more precise. One way to define ``fast" for a parametrized class of graphs (such as the hypercube, where the parameter is the dimension) is to say that the hitting time must be $\mathcal{O}(\log N)$ (exponentially smaller), where $N$ is the number of vertices of the graph. Using this notion, we can expect fast hitting times to exist in graphs which have quotient graphs with an exponentially smaller number of vertices. While this is not necessarily a sufficient condition for fast hitting times, it is interesting to observe that both the quantum walk search algorithm on the hypercube and the glued trees graph are examples of symmetric graphs where the quotient graph is exponentially smaller than the original graph. In the case of the hypercube, the hitting time for the effective walk on the quotient graph is exponentially smaller than the number of vertices in the original graph, and exponentially smaller than the classical hitting time; the same is true of the continuous-time walk on the glued-trees graph.  It is interesting to note that in both of these cases the quotient graph is a finite line. This seems to suggest that if we can identify graphs which have the line as a quotient graph, they may be fruitful ground to look for more examples of walks with fast hitting times.  This remains very much an open question, but it is our belief that graph symmetry is of vital importance in the understanding of hitting times for quantum walks, and that understanding the structure of quotient graphs is the key to further progress.

\section*{Acknowledgments}

We would like to thank Igor Devetak, Viv Kendon, Martin Varbanov and Jason Fulman for helpful conversations.  This work was supported in part by NSF Grant No. EMT-0524822 and NSF CAREER Grant No. 0448658.

\end{document}